\documentclass[preprint,preprintnumbers,amsmath,amssymb]{revtex4}
\usepackage{amsfonts}    
\usepackage{amssymb}
\usepackage{latexsym}
\usepackage{eepic}
\usepackage{graphicx}
\usepackage[usenames,dvipsnames]{color}

\newcommand{\al}{\alpha}
\newcommand{\pa}{\partial}
\newcommand{\ep}{\epsilon}

\newcommand{\la}{\lambda}

\newcommand{\Om}{\Omega}
\newcommand{\om}{\omega}
\newcommand{\de}{\delta}

\newcommand{\De}{\Delta}
\newcommand{\vphi}{\varphi}

\newcommand{\tha}{\theta}

\newcommand{\rar}{\rightarrow}

\newcommand{\non}{\nonumber}

\begin{document}

\title{The quantum $H_3$ integrable system}

\author{Marcos A. G. Garc\'ia}
\email{alejandro.garcia@nucleares.unam.mx}
\affiliation{Instituto de Ciencias Nucleares,
      Universidad Nacional Aut\'onoma de M\'exico,
      Apartado Postal 70-543, 04510 M\'exico, D.F., Mexico}

\author{Alexander V. Turbiner}
\email{turbiner@nucleares.unam.mx}
\affiliation{Instituto de Ciencias Nucleares,
    Universidad Nacional Aut\'onoma de M\'exico,
    Apartado Postal 70-543, 04510 M\'exico, D.F., Mexico}

\date{June 28, 2010}

\begin{abstract}
The quantum $H_3$ integrable system is a $3D$ system with rational
potential related to the non-crystallographic root system $H_3$. It
is shown that the gauge-rotated $H_3$ Hamiltonian as well as one of
the integrals, when written in terms of the invariants of the
Coxeter group $H_3$, is in algebraic form: it has polynomial
coefficients in front of derivatives. The Hamiltonian has
infinitely-many finite-dimensional invariant subspaces in
polynomials, they form the infinite flag with the characteristic
vector $\vec \al\ =\ (1,2,3)$. One among possible integrals is found
(of the second order) as well as its algebraic form. A hidden
algebra of the $H_3$ Hamiltonian is determined. It is an
infinite-dimensional, finitely-generated algebra of differential
operators possessing finite-dimensional representations
characterized by a generalized Gauss decomposition property. A
quasi-exactly-solvable integrable generalization of the model is
obtained. A discrete integrable model on the uniform lattice in a
space of $H_3$-invariants "polynomially"-isospectral to the quantum
$H_3$ model is defined.
\end{abstract}

\maketitle

\section{Introduction}

About 30 years ago, Olshanetsky and Perelomov developed the
Hamiltonian Reduction Method, later known as the Projection Method
(for a review, see \cite{Olshanetsky:1983}). This method provides an
opportunity to construct on a regular basis the non-trivial
multidimensional quantum (and classical) Hamiltonians, which are
associated to the crystallographic root spaces of the classical
($A_N, B_N, C_N, D_N$) and exceptional ($G_2, F_4, E_{6,7,8}$) Lie
algebras. All these systems are symmetric with respect to the
corresponding Weyl group transformations. The Olshanetsky--Perelomov
Hamiltonians have the property of complete integrability (the number
of integrals of motion in involution is equal to the dimension of
the configuration space). There are three types of the Hamiltonians
with rational, trigonometric and elliptic potentials, respectively.
The Hamiltonians with rational and trigonometric potentials are
exactly solvable (the spectrum can be found explicitly, in a form of
a first- or second-degree polynomial in the quantum numbers,
respectively).

A Hamiltonian with rational potential associated to a Lie algebra $g$ of
rank $N$ with root space $\De$ has a form
\begin{equation}
\label{H}
{\cal H}_\Delta = \frac{1}{2}\sum_{k=1}^{N}
\bigg[-\frac{\pa^{2}}{\pa x_{k}^{2}}+\om^{2}x_{k}^{2}\bigg] +
\frac{1}{2}\sum_{\al\in\mathcal{R}^{+}} g_{|\al|}
\frac{|\al|^{2}}{(\al\cdot x)^{2}}\ ,
\end{equation}
where $R_+ \in \Delta$ is the set of positive roots, $\om\in
\mathbb{R}^+$ is a real parameter,
$g_{|\al|}=\nu_{|\al|}(\nu_{|\al|}-1)$ are coupling constants
depending only on  the root length, and $x = (x_1, x_2,\ldots,x_N)$
is the coordinate vector. The configuration space is the principal
Weyl chamber of the root space (see \cite{Olshanetsky:1983}). The
ground state eigenfunction and its eigenvalue are given by
\begin{equation}
\label{Psi_0}
    \Psi_0 (x) \ =\ \Big( \prod_{\al\in
    R_+} (\al\cdot x)^{\nu_{|\al|}} \Big)
    \exp\left(-\frac{\om}{2}t_2^{(\Omega)}\right)
\end{equation}
and its eigenvalue
\begin{equation}
\label{E_0}
    E_0 = \bigg(\frac{N}{2}+\sum_{\al\in\mathcal{R}^+}\nu_{\
    |\al|}\bigg) \om \ ,
\end{equation}
where $t^{(\Om)}_2$ is the invariant of the degree two (for
definition see below). It is indicated in \cite{Olshanetsky:1983}
that the Hamiltonian (\ref{H}) with the property (2)-(3)
can be introduced for the non-crystallographic root systems $H_3$, $H_4$
and dihedral $I_2(m)$ with a Coxeter group as a symmetry of the system.
All that is not true for the $H_3$, $H_4$ and $I_2(m)$ Hamiltonians with
trigonometric or elliptic potential. The complete integrability of (1)
for the case of non-crystallographic root systems has been proven in
\cite{Sasaki:2000} using the formalism of quantum Lax pairs.
\smallskip

Following \cite{Turbiner:2005_1}, we make three definitions.

\medskip

{\sc Definition 1.} A multivariate linear differential operator is
said to be in algebraic form if its coefficients are polynomials in
the independent variable(s). It is called algebraic if by an
appropriate change of the independent variable(s), it can be written
in an algebraic form.

\medskip

{\sc Definition 2}. Consider a finite-dimensional (linear) space of multivariate
polynomials defined as a linear space spanned in the following way:
\[
 { P}^{(\vec \al)}_n \ = \ \langle x_1^{p_1}
x_2^{p_2} \ldots x_d^{p_d} | 0 \leq \al_1 p_1 + \al_2 p_2 +\ldots +
\al_d p_d \leq n \rangle\ \ ,
\]
where the $\al$'s are positive integers  and $n\in \mathbb{N} $. It
represents the Newton polytope in a form of a rectangular pyramid.
Its {\it characteristic vector} is the \hbox{$d$-dimensional} vector
with components $\al_i$:
\begin{equation}
 \vec \al = (\al_1, \al_2, \ldots \al_d)\ .
\end{equation}
For some characteristic vectors, the corresponding polynomial spaces may have a
Lie-algebraic interpretation, in that they are the finite-dimensional representation
spaces for some Lie algebra of differential operators. The smallest characteristic vector is $ \vec \al_0 = (1, 1, \ldots, 1)$. It corresponds to the finite-dimensional representation space for the Lie algebra $gl(d+1)$ of the first order differential operators acting in ${\mathbf R}^d$ \cite{Turbiner:1994}. We will call such a space the {\it basic} space as well as the associated flag will be called the {\it basic} flag.

\medskip

{\sc Definition 3.} Take the infinite set of spaces of multivariate
polynomials $P_n\equiv {P}^{(\vec \al)}_{n}$, $n \in \mathbb{N}$,
defined as above, and order them by inclusion:
\[
{P }_0 \subset  { P}_1 \subset {P}_2 \subset \ldots
 \subset  {P}_n  \subset \ldots \ .
\]
Such an object is called an {\em infinite flag (or filtration)}, and
is denoted ${P}^{(\vec \al)}$. If a linear differential operator
preserves such an infinite flag, it is said to be  {\it
exactly-solvable}. It is evident that every such  operator is
algebraic (see \cite{Turbiner:1994}). If the spaces $P_n$ can be
viewed as the finite-dimensional representation spaces of some (Lie)
algebra $g$, then $g$ is called the {\em hidden algebra} of the
exactly-solvable operator.

If a linear operator preserves several flags and among them there is
a flag for which ${\dim} P_n$ is {\em maximal} for any given $n$,
such a flag is called {\em minimal}. Every flag can be characterized
by a normal vector ${\bf n}_{\vec \al}$ to the hyperplane $\al_1 p_1
+ \al_2 p_2 + \ldots + \al_d p_d = n$ of the base of the Newton
polytope. This normal vector is, in fact, the characteristic vector.
It is clear that for minimal characteristic vector the angle with
basic characteristic vector $(\widehat{{\vec \al},{\vec \al}_0})$ is
minimal.

For any root system $\De$ there exist $N=$ rank$(\De)$ homogeneous,
algebraically independent polynomials which are invariant with
respect to the Coxeter group. They are called the invariants.
The lowest possible degrees $a$ of these invariants are the {\em degrees}
of the Coxeter group.
Each invariant is defined ambiguously, up to a non-linear
combination of the invariants of the lower degrees.

One of the ways to find an invariant of degree $a$ (denoted as
$t_a^{(\Om)}$) is to make averaging over an orbit $\Om$,
\begin{equation}
\label{pol_inv}
    t_a^{(\Om)}(x)\ =\ \sum_{w \in\ \Om}(w \cdot x)^{a}\ ,
\end{equation}
(see e.g. \cite{Bourbaki}), where $x$'s are some formal variables
which can be identified with the Cartesian coordinates. It is worth
mentioning that for any Coxeter group there exists a second degree
invariant $t_2^{(\Om)}$, this invariant does not depend on the chosen
orbit. Later on we will use the invariants (\ref{pol_inv}) as new
variables. We will call them the {\em orbit variables}.

For all the crystallographic root systems algebraic representations
of all quantum Hamiltonians, both rational and trigonometric,
have been found (\cite{Ruhl:1995}-\cite{Turbiner:2010}). The general
strategy which was used to find a minimal flag for the rational Hamiltonians
is the following: (i) as a first step we consider the
similarity-transformed version of (\ref{H}), namely $h \propto
\Psi_0^{-1} ({\cal H} - E_0) \Psi_0$, (ii) then, we choose a certain orbit to
construct a particular set of variables which lead to an algebraic
form of the transformed Hamiltonian $h$, (ii) finally, exploiting the
ambiguity in the definition of polynomial invariants of the fixed
degrees we search for variables for which the flag of invariant
subspaces of (1) is minimal flag. A primary goal of
this paper is to show that the same strategy can be applied for a study
of the rational $H_3$ Hamiltonian (1). We find the algebraic form of the rational
$H_3$ Hamiltonian and a minimal flag of its invariant subspaces.
Furthermore, we demonstrate that any (invariant) subspace from the
minimal flag is a representation space of a finite-dimensional
representation of a certain infinite-dimensional, finitely-generated algebra.
This algebra is the hidden algebra of the $H_3$ system. A similar analysis
is done for one of the integrals of the $H_3$ system.

Another goal of the paper is to find a quasi-exactly-solvable
generalization of the $H_3$ rational model. By definition a linear
differential operator is quasi-exactly-solvable (QES) if it
preserves a finite-dimensional functional space with an explicitly
indicated basis (see e.g. \cite{Turbiner:1988}). Thus, it implies
that the QES operator has a finite-dimensional invariant subspace
spanned by known functions. Furthermore, one can indicate explicitly
a basis where the operator being written in the matrix form has a
block-triangular form. In practice, for all known examples of the
QES operators the finite-dimensional invariant subspace is a space
of inhomogeneous polynomials in one or several variables. In many
cases the space of polynomials can be identified with a finite
dimensional representation space of a Lie algebra of differential
operators of the first order. In the case of crystallographic root
systems a certain QES generalization has been found for each
particular rational Hamiltonian \cite{Turbiner:2005_2}. All those
examples are related with the existence of the hidden $sl(2)$
algebra. We show that a similar $sl(2)$-quasi-exactly-solvable
generalization of the $H_3$ model exists.

Finally, we show that in the $H_3$ orbit space (in the space of the
$H_3$ invariants $t_a^{(\Om)}$) there exists a discrete model
defined on three-dimensional uniform lattice with polynomial
eigenfunctions which is isospectral to the rational $H_3$ model and
integrable. We will call this model the "discrete $H_3$ rational
model".

\medskip

\section{The Hamiltonian}

The Hamiltonian of the rational $H_3$ model (see (\ref{H})) is
invariant wrt the $H_3$ Coxeter group, which is the full symmetry
group of the icosahedron.
This discrete group is subgroup of $O(3)$ and its dimension is 120 
(see e.g. \cite{Grove:1985}). In the Cartesian
coordinates $x_1,x_2,x_3$ the Hamiltonian has the form
\begin{equation}
\label{H_H3}
\begin{aligned}
    {\cal H}_{H_3} = \ &\frac{1}{2}\sum_{k=1}^{3}\left[-\frac{\pa^{2}}{\pa
    x_{k}^{2}}+\om^{2}x_{k}^{2}+\frac{g}{x_{k}^{2}}\right]\\
    &+\sum_{\{i,j,k\}}\,\sum_{\mu_{1,2}=0,1}\frac{2g}{[x_{i}+(-1)^{\mu_1}
    \varphi_{+}x_{j}+(-1)^{\mu_2}\varphi_{-}x_{k}]^{2}}\,,\quad x \in {\mathbf R}^3
\end{aligned}
\end{equation}
where $\{i,j,k\}=\{1,2,3\}$ and its even permutations. Here
$g=\nu(\nu-1)>-1/4$ is the coupling constant,
$\varphi_{\pm}=(1\pm\sqrt{5})/2$ the {\em golden ratio} and its
algebraic conjugate. We choose as the configuration space the
fundamental domain of the $H_3$ group -- the space bounded by three
planes
\begin{equation}
\label{confspace}
x_1=0\,,\quad x_3=0 \quad\text{and}\quad
x_3+\varphi_+x_1+\varphi_-x_2=0
\end{equation}
for $x_1,x_2,x_3 \geq 0$.

The ground state eigenfunction and its eigenvalue are
\begin{equation}
\label{Psi_H3}
 \Psi_{0}(x)=\Delta_{1}^{\nu}\Delta_{2}^{\nu}
 \exp \bigg(-\frac{\om}{2}\sum_{k=1}^{3}x_{k}^{2} \bigg)\ ,
 \quad E_{0}=\frac{3}{2}\om(1+10\nu)\ ,
\end{equation}
where
\begin{eqnarray}
\label{D12}
 \De_{1}&=&\prod_{k=1}^{3}x_{k}\ ,\\
 \De_{2}&=&\prod_{\{i,j,k\}}\,\prod_{\mu's=0,1}[x_{i}+(-1)^{\mu_1}
 \varphi_{+}x_{j}+(-1)^{\mu_2}\varphi_{-}x_{k}]\,.
\end{eqnarray}
The ground state eigenfunction (\ref{Psi_H3}) does not vanish in the
configuration space (\ref{confspace}).

The main object of our study is the gauge-rotated Hamiltonian
(\ref{H_H3}) with the ground state eigenfunction (\ref{Psi_H3})
taken as a factor,
\begin{equation}
\label{h_H3}
 h_{\rm H_3} \ =\ -2(\Psi_{0})^{-1}({\cal H}_{\rm H_3}-E_0)(\Psi_{0}) \ ,
\end{equation}
where $E_0$ is given by (\ref{Psi_H3}). The gauge rotated operator (\ref{h_H3})
is the second-order differential operator without free term. By construction
its lowest eigenfunction is a constant and the lowest eigenvalue is equal to
zero. Now let us introduce new variables in (\ref{h_H3}).

The $H_3$ root space is characterized by three fundamental weights
$w_c,\ c=1,2,3$ (see e.g. \cite{Humphreys:1990}). Taking action of
all group elements on fundamental weight $\om_c$ we generate orbit
$\Om_c$ of a certain length (length $\equiv$ \#elements of the
orbit). The results are summarized as
\begin{center}
\begin{tabular}{cc}
\hline
\rule[-8pt]{0pt}{22pt}{weight}    & \quad orbit length\\
\hline
\rule[-8pt]{0pt}{22pt}{$w_1=(0,\ \varphi_{+},\ 1)$}     & 12 \\
\rule[-8pt]{0pt}{22pt}{$w_2=(1,\ \varphi_{+}^{2},\ 0)$} & 20 \\
\rule[-8pt]{0pt}{22pt}{$w_3=(0,\ 2\varphi_{+},\ 0)$}    & 30 \\
\hline\\
\end{tabular}
\end{center}
In order to find $H_3$-invariants (\ref{pol_inv}) we choose for simplicity
the shortest orbit $\Om (w_1)$ and make averaging,
\begin{equation}
\label{invvars}
    t_{a}^{(\Om)}(x)=\sum_{w \in\ \Omega(w_1)}(w \cdot x)^{a}\ ,
\end{equation}
where $a=2,6,10$ are the degrees of the $H_3$ group. These invariants
are defined ambiguously, up to a non-linear combination of the invariants
of the lower degrees
\begin{equation}
\label{coords}
\begin{aligned}
t_{2}^{(\Om)}&\mapsto t_{2}^{(\Om)}\,,\\
t_{6}^{(\Om)}&\mapsto t_{6}^{(\Om)} + A \ (t_{2}^{(\Om)})^{3}\,,\\
t_{10}^{(\Om)}&\mapsto t_{10}^{(\Om)} + B\
(t_{2}^{(\Om)})^{2}t_{6}^{(\Om)} + C \ (t_{2}^{(\Om)})^{5}\,,
\end{aligned}
\end{equation}
where $A,B,C$ are parameters. Now we can make a change of variables
in the gauge-rotated Hamiltonian (\ref{h_H3}):
$$(x_1,x_2,x_3) \rar (t_{2}^{(\Om)},t_{6}^{(\Om)},t_{10}^{(\Om)})\ .$$
The first observation is that the transformed Hamiltonian $h_{\rm H_3}(t)$
(\ref{h_H3}) takes on an algebraic form for any value of the parameters
$A,B,C$ in variables $t$'s (\ref{coords}). The second observation
is that for any value of the parameters $A,B,C$ there exists a
flag of invariant subspaces in polynomials of the Hamiltonian $h_{\rm H_3}(t)$.
Our goal is to find the parameters for which $h_{\rm H_3}(t)$ preserves the
minimal flag. After some analysis we found such a set of parameters
\begin{equation}
A\ =\ -\frac{13}{10}\,,\quad B\ =\ -\frac{76}{15}\,,\quad C\ =\ \frac{1531}{375}\ .
\end{equation}
The $t$-variables (\ref{coords}) for such values of parameters,
which we denote as $\tau$-variables, are
\begin{equation}
\label{tau}
\begin{aligned}
\tau_1=\,&x_1^2+x_2^2+x_3^2\,,\\
\tau_2=\,&-\frac{3}{10}\,(x_1^6+x_2^6+x_3^6)+\frac{3}{10}(2-5\vphi_+)\,(x_1^2x_2^4+x_2^2x_3^4+x_3^2x_1^4)\\
&+\frac{3}{10}(2-5\vphi_-)\,(x_1^2x_3^4+x_2^2x_1^4+x_3^2x_2^4)-\frac{39}{5}\,(x_1^2x_2^2x_3^2)\,,\\
\tau_3=\,&\frac{2}{125}\,(x_1^{10}+x_2^{10}+x_3^{10})+\frac{2}{25}(1+5\vphi_-)\,(x_1^8x_2^2+x_2^8x_3^2+x_3^8x_1^2)\\
&+\frac{2}{25}(1+5\vphi_+)\,(x_1^8x_3^2+x_2^8x_1^2+x_3^8x_2^2)+\frac{4}{25}(1-5\vphi_-)\,(x_1^6x_2^4+x_2^6x_3^4+x_3^6x_1^4)\\
&+\frac{4}{25}(1-5\vphi_+)\,(x_1^6x_3^4+x_2^6x_1^4+x_3^6x_2^4)-\frac{112}{25}\,(x_1^6x_2^2x_3^2+x_2^6x_3^2x_1^2+x_3^6x_1^2x_2^2)\\
&+\frac{212}{25}\,(x_1^2x_2^4x_3^4+x_2^2x_3^4x_1^4+x_3^2x_1^4x_2^4)\,.
\end{aligned}
\end{equation}
The Hamiltonian $h_{\rm H_3}(\tau)$ has infinitely-many finite-dimensional invariant subspaces
\begin{equation}
\label{minflag}
 \mathcal{P}_{n}^{(1,2,3)}=\langle
 \tau_1^{p_1}\tau_2^{p_2}\tau_3^{p_3}\,|
 \,0\leq p_1+2p_2+3p_3\leq n\rangle\ , \quad n=0,1,2,\ldots \ .
\end{equation}
which form the minimal flag. Its characteristic vector is
\begin{equation}
\label{min-ch-vec}
   \vec{\al}_{min}\ =\ (1,2,3)\ .
\end{equation}
It is worth noting that each particular space $\mathcal{P}_{n}^{(1,2,3)}$ (\ref{minflag}) as well as the whole flag are invariant with respect to a weighted projective transformation
\begin{eqnarray}
\label{wpt}
\notag \tau_1&\rightarrow&\tau_1+a\,,\\
 \tau_2&\rightarrow&\tau_2+b_1\,\tau_1^2+b_2\,\tau_1+b_3\,,\\
\notag
 \tau_3&\rightarrow&\tau_3+c_1\,\tau_1\tau_2+c_2\,\tau_1^3
 +c_3\,\tau_2+c_4\,\tau_1^2+c_5\,\tau_1+c_6\,.
\end{eqnarray}
where $\{a,b,c\}$ are parameters. It manifests a hidden invariance of the Hamiltonian (\ref{H_H3}). It is seen in a clear way in the space of orbits only.

Finally, the gauge-rotated Hamiltonian (\ref{h_H3}) in the $\tau$-coordinates is written as
\begin{equation}
\label{h_H3_tau}
 {h}_{\rm H_3}\ =\
 \sum_{i,j=1}^{3} {A}_{ij}({\tau})
 \frac{\pa^2}{\pa {{\tau}_i} \pa {{\tau}_j} } +
 \sum_{i=1}^{3}{B}_i({\tau}) \frac{\pa}{\pa {\tau}_i} \ ,\ {A}_{ij}={A}_{ji}
 \ ,
\end{equation}
with the coefficient functions
\begin{eqnarray}
\label{A-B}
A_{11}&=&4\tau_{1}\,,\non \\
A_{12}&=&12\tau_{2}\,,\non \\
A_{13}&=&20\tau_{3}\,,\non \\
A_{22}&=&-\frac{48}{5}\tau_{1}^2\tau_2+\frac{45}{2}\tau_3\,,\non \\
A_{23}&=&\frac{16}{15}\tau_1\tau_2^2-24\tau_1^2\tau_3\,,\non \\
A_{33}&=&-\frac{64}{3}\tau_1\tau_2\tau_3+\frac{128}{45}\tau_2^3\,,
\\
B_{1}&=&6(1+10\nu) - 4\om\tau_1\,,\non \\
B_{2}&=&-\frac{48}{5}(1+5\nu)\tau_{1}^{2}-12\om\tau_2\,,\non \\
B_{3}&=&-\frac{64}{15}(2+5\nu)\tau_1\tau_2-20\om\tau_3\ .\non
\end{eqnarray}
It can be easily checked that the operator (\ref{h_H3_tau}) is triangular
with respect to action on monomials $\tau_1^{p_1}\tau_2^{p_2}\tau_3^{p_3}$.
One can find the spectrum of (\ref{h_H3_tau}) $h_{\rm H_3}
\varphi=-2\epsilon\varphi$ explicitly
\begin{equation}\label{spectrumh}
\epsilon_{n_{1},n_{2},n_{3}} = 2\om(n_{1}+3n_{2}+5n_{3})\,,
\end{equation}
where $n_i=0,1,2,\ldots$. Degeneracy of the spectrum is
related to the number of solutions of the equation $n_1+3n_2+5n_3=k$
for $k=0,1,2\ldots$ in non-negative numbers $n_{1,2,3}$. The spectrum $\ep$ does not
depend on the coupling constant $g$ and it is equidistant. It coincides
to the spectrum of $3D$ anisotropic harmonic oscillator with
frequencies $(2\om,6\om,10\om)$. The energies of the original rational
$H_3$ Hamiltonian (\ref{H_H3}) are $E=E_0+\ep$.

The boundary of the configuration space of the rational $H_3$ model
(\ref{H_H3}) in the $\tau$ variables is determined by the zeros of
the ground state eigenfunction, hence, by pre-exponential factor in
(\ref{Psi_H3}). It is the algebraic surface of degree 15 in Cartesian coordinates being a product of monomials. In $\tau$-coordinates it can be written as
\begin{equation}
\label{confspacet}
 12960\tau_1^5\tau_3^2 - 5760\tau_1^4\tau_2^2\tau_3 +
 640\tau_1^3\tau_2^4 + 54000\tau_1^2\tau_2\tau_3^2 -21600\tau_1\tau_2^3\tau_3 + 2304\tau_2^5 + 50625\tau_3^3\ =\ 0\ ,
\end{equation}
which is the algebraic surface of degree seven; the equation
contains monomials of the degrees 7, 5 and 3. It is worth mentioning
that l.h.s. of (\ref{confspacet}) is proportional to the square of
Jacobian, $ J^2 (\frac{\pa \tau}{\pa x}) $.

\medskip

\section{Integral}

The Hamiltonian (\ref{H_H3}) being written in spherical
coordinates $(r, \tha, \phi)$ takes a very simple form
\begin{equation}
\label{H_H_3_sphe}
{\cal H}_{H_3}=-\frac{1}{2}\De^{(3)} + \frac{1}{2}\om^2 r^2 +
\frac{W(\theta,\phi)}{r^2}\ ,
\end{equation}
where $\De^{(3)}$ is the $3D$ Laplacian and the angular function
\begin{equation}
\begin{aligned}
W(\theta,\phi) &=\frac{2\nu(\nu-1)}{(s_{\tha}c_{\phi}+\varphi_+s_{\tha}s_{\phi}+\varphi_-c_{\tha})^2}
+\frac{2\nu(\nu-1)}{(s_{\tha}c_{\phi}-\varphi_+s_{\tha}s_{\phi}+\varphi_-c_{\tha})^2}\\
&+\frac{2\nu(\nu-1)}{(s_{\tha}c_{\phi}+\varphi_+s_{\tha}s_{\phi}-\varphi_-c_{\tha})^2}
+\frac{2\nu(\nu-1)}{(s_{\tha}c_{\phi}-\varphi_+s_{\tha}s_{\phi}-\varphi_-c_{\tha})^2}\\
&+\frac{2\nu(\nu-1)}{(s_{\tha}s_{\phi}+\varphi_+c_{\tha}+\varphi_-s_{\tha}c_{\phi})^2}
+\frac{2\nu(\nu-1)}{(s_{\tha}s_{\phi}-\varphi_+c_{\tha}+\varphi_-s_{\tha}c_{\phi})^2}\\
&+\frac{2\nu(\nu-1)}{(s_{\tha}s_{\phi}+\varphi_+c_{\tha}-\varphi_-s_{\tha}c_{\phi})^2}
+\frac{2\nu(\nu-1)}{(s_{\tha}s_{\phi}-\varphi_+c_{\tha}-\varphi_-s_{\tha}c_{\phi})^2}\\
&+\frac{2\nu(\nu-1)}{(c_{\tha}+\varphi_+s_{\tha}c_{\phi}+\varphi_-s_{\tha}s_{\phi})^2}
+\frac{2\nu(\nu-1)}{(c_{\tha}-\varphi_+s_{\tha}c_{\phi}+\varphi_-s_{\tha}s_{\phi})^2}\\
&+\frac{2\nu(\nu-1)}{(c_{\tha}+\varphi_+s_{\tha}c_{\phi}-\varphi_-s_{\tha}s_{\phi})^2}
+\frac{2\nu(\nu-1)}{(c_{\tha}-\varphi_+s_{\tha}c_{\phi}-\varphi_-s_{\tha}s_{\phi})^2}\\
&+\frac{\nu(\nu-1)}{2s_{\tha}^2c_{\phi}^2}+\frac{\nu(\nu-1)}{2s_{\tha}^2s_{\phi}^2}+
\frac{\nu(\nu-1)}{2c_{\phi}^2}\ .
\end{aligned}
\end{equation}
Here, for the sake of simplicity we denoted $c_{\vartheta}\equiv\cos \vartheta$,
$s_{\vartheta}\equiv\sin \vartheta$. It is seen immediately, that the Schroedinger equation (\ref{H_H_3_sphe}) admits a separation of radial variable $r$:
any solution can be written in factorized form
\begin{equation}
\Psi(r,\tha,\phi)=R(r)Q(\tha,\phi)\ .
\end{equation}
Functions $R$ and $Q$ are the solutions of the
equations
\begin{equation}
\bigg[-\frac{1}{2 r^2}\frac{\pa}{\pa r}\bigg(r^2\frac{\pa}
{\pa r}\bigg) + \frac{1}{2}\om^2r^2 + \frac{\gamma}{r^2}\bigg]R(r)
\ =\ E R(r)\ ,
\end{equation}
\begin{equation}
\label{eigf}
\mathcal{F}\ Q(\theta,\phi)=\gamma\ Q(\theta,\phi)\ ,
\end{equation}
respectively, while $\gamma$ is the constant of separation. The operator
$\mathcal{F}$ has the form
\begin{equation}
\label{opf}
\mathcal{F}\ =\ \frac{1}{2}\ \mathcal{L}^2 + W(\theta,\phi)\ ,
\end{equation}
where $\mathcal{L}$ is the angular momentum operator:
\begin{equation*}
\mathcal{L}^2=-\frac{1}{\sin\theta}\frac{\pa}{\pa\theta}\bigg(\sin\theta
\frac{\pa}{\pa\theta}\bigg) - \frac{1}{\sin^2\theta}\frac{\pa^2}{\pa\phi^2}\ .
\end{equation*}
It can be immediately checked that the Hamiltonian ${\cal H}_{H_3}$ and $\mathcal{F}$
commute,
\begin{equation}
[{\cal H}_{H_3},\mathcal{F}]\ =\ 0\ .
\end{equation}
Hence, $\mathcal{F}$ is an integral of motion. Thus, it has common eigenfunctions with the Hamiltonian ${\cal H}_{H_3}$.

Let us make a gauge rotation of the operator $\mathcal{F}$
(\ref{opf}) with the ground state function $\Psi_0$ as a gauge factor,
\begin{equation}
 f\ =\ (\Psi_0)^{-1}(\mathcal{F}-\gamma_{0})\Psi_0\ ,\qquad
\gamma_0=\frac{15}{2}\nu(1+15\nu)\ ,
\end{equation}
where $\gamma_0$ is the lowest eigenvalue of $\mathcal{F}$, and make a change of variables to the $\tau$ variables (\ref{tau}). The operator $f$ has an algebraic form,
\begin{equation}
\label{fmin}
f\ =\ \sum_{i,j=1}^{3}F_{ij}\frac{\pa^{2}}{\pa\tau_{i}\pa\tau_{j}}
 +\sum_{j=1}^{3}G_{j}\frac{\pa}{\pa\tau_{j}} \ ,\quad {F}_{ij}={F}_{ji}
\end{equation}
where
\begin{eqnarray}
F_{11}&=& 0\,,\non\\
F_{12}&=& 0\,,\non\\
F_{13}&=& 0\,,\non\\
F_{22}&=&\frac{24}{5}\tau_1^3\tau_2-\frac{45}{4}\tau_1\tau_3+18\tau_2^2\,,\non \\
F_{23}&=&-\frac{8}{15}\tau_1^2\tau_2^2+12\tau_1^3\tau_3+30\tau_2\tau_3\,, \\
F_{33}&=&-\frac{64}{45}\tau_1\tau_2^3+\frac{32}{3}\tau_1^2\tau_2\tau_3+50\tau_3^2\,,\non\\
G_{1}&=& 0\,,\non\\
G_{2}&=&\frac{24}{5}(1+5\nu)\tau_1^3+3(7+30\nu)\tau_2\,,\non \\
G_{3}&=&\frac{32}{15}(2+5\nu)\tau_1^2\tau_2+5(11+30\nu)\tau_3\ .\non
\end{eqnarray}
It is worth noting that in the operator $f$ the variable $\tau_1$
appears as a parameter. It implies that any eigenfunction of
$h_{H_3}$ which depends on $\tau_1$ only (see below Ch.V for a
discussion) is the eigenfunction of the integral $f$ with zero
eigenvalue.

It can be also shown that the operator $f$ has infinitely many finite-dimensional invariant subspaces in polynomials
\begin{equation}
 \mathcal{P}_{n}^{(1,3,5)}\ =\
 \langle \tau_1^{p_1}\tau_2^{p_2}\tau_3^{p_3}\,|
 \,0\leq p_1+3p_2+5p_3\leq n\rangle\ ,\ n=0,1,2, \ldots \ .
\end{equation}
which form a flag with characteristic vector $(1,3,5)$.

The spectrum of the integral $\mathcal{F}$ can be found in a closed
form,
\begin{equation}
\label{gamma}
\gamma_{0, k_2, k_3}\ =\
 2(3k_2+5k_3)^2-30k_2k_3+(1+30\nu)(3k_2+5k_3)+\gamma_0\ ,
\end{equation}
where $k_2,k_3=0,1,2,\ldots$ and $\gamma_0$ is given by (29).

It can be shown that the Hamiltonian $h_{H_3}$ has a certain
degeneracy -- it preserves two different flags: one with minimal
characteristic vector (1,2,3) and another one with characteristic
vector (1,3,5). The fact that the operator $h_{H_3}$ with
coefficients (\ref{A-B}) commutes with $f$ given by (\ref{fmin})
implies that common eigenfunctions of the operators $h_{H_3}$ and
$f$ are elements of the flag of spaces $\mathcal{P}^{(1,3,5)}$.

Let us denote $\phi_{n,i}$ the common eigenfunctions of $h_{H_3}$ and
$f$ which are elements of the invariant space $P^{(1,3,5)}_n$ and their respectful eigenvalues $\ep_{n,i}, \gamma_{n,i}$.
The index $i$ numerates these eigenfunction for given $n$ starting from 0.
The function $\phi_{n,i}$ is related to the eigenfunction
of the Hamiltonian $\mathcal{H}_{H_3}$ (\ref{H_H3}) (and the integral ${\cal F}$) through $\Psi_{n,i}=\Psi_{0}\phi_{n,i}$. Thus, the eigenfunctions $\{\phi\}$ are
orthogonal with the weight factor $|\Psi_0|^2$. As an illustration let us
give explicit expressions for several eigenfunctions $\phi_{n,i}$ and their respectful eigenvalues,

\begin{itemize}
\item $n=0$
\[
\phi_{0,0}= 1\ ,\quad \ep_{0,0}= 0\ ,\quad \gamma_{0,0}= 0\ .
\]
\item $n=1$
\[
\phi_{1,0}= \tau_1-\frac{3}{2\om}(1+10\nu)\ ,\quad \epsilon_{1,0}=
2\om\ ,\quad \gamma_{1,0}= 0\ .
\]
\item $n=2$
\[
\phi_{2,0}=
\tau_1^2-\frac{5}{\om}(1+6\nu)\tau_1+\frac{15}{4\om^2}(1+6\nu)(1+10\nu)\
,\quad \epsilon_{2,0}= 4\om\ ,\quad \gamma_{2,0}= 0\ .
\]
\item $n=3$
\begin{equation*}
\begin{aligned}
\phi_{3,0}=&\tau_1^3-\frac{3}{2\om}(7+30\nu)\tau_1^2+\frac{15}{4\om^2}(1+6\nu)(7+30\nu)\tau_1-\frac{15}{8\om^3}(1+6\nu)(7+30\nu)(1+10\nu)\,,\\
\epsilon_{3,0}&=6\om\ ,\quad \gamma_{3,0}= 0\ .\\[10pt]
\phi_{3,1}=&\tau_2+\frac{8(1+5\nu)}{5(7+30\nu)}\tau_1^3\ ,\quad
\epsilon_{3,1}=6\om\ ,\quad \gamma_{3,1}=21+90\nu\ .
\end{aligned}
\end{equation*}
\end{itemize}

As stated before, the Hamiltonian $h_{H_3}$ preserves two flags with characteristic vectors $(1,2,3)$ and $(1,3,5)$, respectively. The angle between the normal vectors of the minimal flag $(1,2,3)$ and of the basic one $(1,1,1)$ is given
\[
\cos\theta_h\ =\ \frac{6}{\sqrt{42}}\quad\text{or}\quad\theta_h\simeq 0.39\ ,
\]
while between the vectors $(1,3,5)$ and $(1,1,1)$
\[
\cos\theta_f\ =\ \frac{9}{\sqrt{105}}\quad\text{or}\quad\theta_f\simeq 0.50\ .
\]

It seems evident that if one or more extra integrals exist they will
take an algebraic form in $\tau$-variables after the gauge rotation.

\medskip

\section{Discrete uniform $H_3$ system}

The existence of the algebraic form of the $H_3$ Hamiltonian in the
space of invariants allows us to construct a discrete system with a
remarkable property of isospectrality. This construction is based on
employment of a quantum canonical transformation as a basis to
perform a discretization of a continuous system \cite{ST:1995}. Such
a procedure was called a {\it Lie-algebraic discretization}. It was
already used in the past to construct the isospectral discrete model
of the harmonic oscillator (the $A_1$ system in the Hamiltonian
Reduction nomenclature) in the space of $\tau=x^2$
\cite{Turbiner:2005}.

Let us introduce a set of the finite-difference operators
\begin{equation}
\label{findifops}
\begin{aligned}
\mathcal{D}_i^{(\de_i)}\
f(\tau_i)&\equiv \frac{f(\tau_i+\de_i)-f(\tau_i)}{\de_i}\
=\ \frac{(e^{\de_i\pa_i}-1)}{\de_i}f(x_i)\
,\\
\mathcal{X}_i^{(\de_i)}\ f(\tau_i)&\equiv \tau_i f(\tau_i-\de_i)\ =\
(\tau_i e^{-\de_i\pa_i}) f(\tau_i)\ ,
\end{aligned}
\end{equation}
where $\de_{i},\ i=1,2,3$ are spacings; here no summation over repeated indexes is implied. The operator $\mathcal{D}_i^{(\de_i)}$ is the finite-difference derivative or discrete momentum; sometimes, it is called the Norlund derivative. The operator $\mathcal{X}_i^{(\de_i)}$ is a discrete analogue of the multiplication operator. The operators $\mathcal{D}_i^{(\de_i)}$ and $\mathcal{X}_i^{(\de_i)}$ form a canonical pair,
\begin{equation}
[\mathcal{D}_i^{(\de_i)},\mathcal{D}_j^{(\de_j)}]=\ 0\ ,\
[\mathcal{X}_i^{(\de_i)},\mathcal{X}_j^{(\de_j)}]=\ 0\ ,\
[\mathcal{D}_i^{(\de_i)},\mathcal{X}_j^{(\de_j)}]=\delta_{ij}\ ,
\end{equation}
for $i,j=1,2,3$.
Hence, the operators (\ref{findifops}) span the 7-dimensional Heisenberg algebra realizing a three-parametric quantum canonical transformation with parameters $\de_{1,2,3}$. In the limit when all $\de_{i}$ tend to zero the operators (\ref{findifops}) gives rise to a standard coordinate-momentum representation,
\[
     {\mathcal{D}}_i^{(\de_i)} \rar \pa_i\ ,\ {\mathcal{X}}_i^{(\de_i)} \rar \tau_i\ .
\]

Take a linear differential operator $\mathcal{L}(\pa_i,\tau_i)$. Consider the eigenvalue problem
\begin{equation}
\label{eigop1}
\mathcal{L}(\pa_i,\tau_i)\ \varphi(\tau)=\la \ \varphi(\tau)\ ,
\end{equation}
and assume it has polynomial eigenfunctions. Performing the canonical
transformation (\ref{findifops}) we arrive at
\begin{equation}
\label{eigop2}
\mathcal{L}(\mathcal{D}_i^{(\de_i)},\mathcal{X}_i^{(\de_i)})\
\varphi(\mathcal{X}_i^{(\de_i)}) |0\rangle\ =\ \la \
\varphi(\mathcal{X}_i^{(\de_i)}) |0\rangle
\end{equation}
In order to make sense to this equation one should introduce the
vacuum $|0\rangle$:
\begin{equation}
\label{vacuum}
\mathcal{D}_i^{(\de_i)}|0\rangle\ =\ 0\ ,\quad i=1,2,3\ .
\end{equation}
Then the equation (\ref{eigop2}) has a meaning of an operator
eigenvalue problem in the Fock space with vacuum (\ref{vacuum}). Now let us show
that the eigenvalue problem (\ref{eigop2}) has polynomial
eigenfunctions and their eigenvalues are the same eigenvalues as for the polynomial
eigenfunctions as the original (continuous) problem (\ref{eigop1}).

In order to exploit the representation (\ref{findifops}) let us
first define the vacuum $|0\rangle$. The condition (\ref{vacuum}) in
explicit form is
\begin{align*}
f(\tau_1+\delta_1,\tau_2,\tau_3)&=f(\tau_1,\tau_2,\tau_3)\ ,\\
f(\tau_1,\tau_2+\delta_2,\tau_3)&=f(\tau_1,\tau_2,\tau_3)\ ,\\
f(\tau_1,\tau_2,\tau_3+\delta_3)&=f(\tau_1,\tau_2,\tau_3)\ .
\end{align*}
Any periodic function with periods $\delta_i$ in the coordinates
$\tau_i$ is the solution of these equations; however, without loss
of generality, we can make the choice
\begin{equation}
\label{vac1}
f(\tau_1,\tau_2,\tau_3)=1\ .
\end{equation}
Let us now define the {\em quasi-monomial}
\begin{equation}
\tau^{(n+1)}=\tau(\tau-\de)(\tau-2\de)\cdots(\tau-n\de)\ .
\end{equation}
Taking into account the relation
\[
\left(\tau
e^{-\de\pa_\tau}\right)^n=\tau^{(n)}e^{-n\de\pa_\tau}
\]
and choosing of vacuum (\ref{vac1}), it is easy to check that
\begin{equation}
\label{dquasi}
\left(\mathcal{X}_i^{(\de_i)}\right)^n |0\rangle=\tau_i^{(n)}\ .
\end{equation}

Now we can relate the solutions of (\ref{eigop2}) with the solutions
of (\ref{eigop1}). Let us assume that
\begin{equation}
\label{sol1}
\varphi(\tau)=\sum\ \alpha_{klm}\ \tau_1^k \tau_2^l \tau_3^m
\end{equation}
is a polynomial solution of the equation (\ref{eigop1}). The
canonical transformation (\ref{findifops}) implies the replacement
of $\tau_i$ by $\mathcal{X}_i^{(\de_i)}$. Taking in account
(\ref{dquasi}) we come to the conclusion that each monomial in
(\ref{sol1}) should be replaced by a quasi-monomial. Hence, the
corresponding polynomial solution of (\ref{eigop2}) is
\begin{equation}
\tilde{\varphi}(\tau)=\sum\ \alpha_{klm}\ \tau_1^{(k)} \tau_2^{(l)}
\tau_3^{(m)}\ ,
\end{equation}
with the same expansion coefficients $\alpha_{klm}$ as in (\ref{sol1})
and the same eigenvalue.

Performing the procedure of canonical discretization (\ref{eigop1})
$\rightarrow$ (\ref{eigop2}) for the $H_3$ Hamiltonian in the algebraic
form $h_{H_3}$ (\ref{h_H3_tau}), we arrive at the following (isospectral)
finite-difference operator:
\begin{equation}
\label{hdiscr}
\tilde{h}_{H_3}\equiv
h_{H_3}(\mathcal{D}_i^{(\de_i)},\mathcal{X}_i^{(\de_i)})\
=\ \sum_{k_1,k_2,k_3} A_{k_1k_2k_3}\
e^{k_1\de_1\pa_1+k_2\de_2\pa_2+k_3\de_3\pa_3}\ ,
\end{equation}
with the following non-vanishing coefficients
\begin{eqnarray}
&A_{0,0,0}&=\quad-\frac{4}{\de_1}(2+\de_1\om)
            \left[\frac{\tau_1}{\de_1}+\frac{3\tau_2}{\de_2}+
            \frac{5\tau_3}{\de_3}\right] -\frac{6}{\de_1}(1+10\nu)\ , \non\\
&A_{1,0,0}&=\quad\frac{2}{\de_1}\left[\frac{2\tau_1}{\de_1}+
   \frac{12\tau_2}{\de_2}+\frac{20\tau_3}{\de_3}+3(1+10\nu)\right]\ , \non\\
&A_{-1,0,0}&=\quad\frac{4}{\de_1^2}(1+\de_1\om)\tau_1\ , \non\\
&A_{-2,0,0}&=\quad\frac{48}{5\de_2}\tau_1(\tau_1-\de_1)
  \left[\frac{2\tau_2}{\de_2}+\frac{5\tau_3}{\de_3}+1+5\nu\right]\ ,\non \\
&A_{0,-1,0}&=\quad\frac{12}{\de_1\de_2}(2+\de_1\om)\tau_2\ ,\non\\
&A_{0,0,-1}&=\quad \frac{5}{2}\left[\frac{8}{\de_1\de_3}(2+\de_1\om)+
                        \frac{9}{\de_2^2}\right]\tau_3\ , \non\\
\end{eqnarray}
\begin{eqnarray}
&A_{0,-3,0}&=\quad\frac{128}{45\de_3^2}\tau_2(\tau_2-\de_2)
                                         (\tau_2-2\de_2)\ , \non\\
&A_{1,-1,0}&=\quad-\frac{24\tau_2}{\de_1\de_2}\ , \non\\
&A_{1,0,-1}&=\quad-\frac{40\tau_3}{\de_1\de_3}\ , \non \\
&A_{-1,-1,0}&=\quad-\frac{32}{15\de_3}\tau_1\tau_2
  \left[\frac{\tau_2}{\de_2}-\frac{20\tau_3}{\de_3}-5(1+2\nu)\right]\ , \non\\
&A_{-1,-2,0}&=\quad\frac{32}{15\de_2\de_3}\tau_1\tau_2(\tau_2-\de_2)\ , \non\\
&A_{-1,-1,1}&=\quad\frac{32}{15\de_3}\tau_1\tau_2\left[\frac{\tau_2}{\de_2}
       -\frac{10\tau_3}{\de_3}-5(1+2\nu)\right]\ , \non\\
&A_{-1,-1,-1}&=\quad-\frac{64}{3\de_3^2}\tau_1\tau_2\tau_3\ , \non\\
&A_{-1,-2,1}&=\quad-\frac{32}{15\de_2\de_3}\tau_1\tau_2(\tau_2-\de_2)\ , \non\\
&A_{-2,1,0}&=\quad-\frac{48}{5\de_2}\tau_1(\tau_1-\de_1)
   \left[\frac{\tau_2}{\de_2}+\frac{5\tau_3}{\de_3}+1+5\nu\right]\ , \non\\
&A_{-2,-1,0}&=\quad-\frac{48}{5\de_2^2}\tau_2\tau_1(\tau_1-\de_1)\ , \non\\
&A_{-2,0,-1}&=\quad-\frac{48}{\de_2\de_3}\tau_3\tau_1(\tau_1-\de_1)\ , \non\\
&A_{-2,1,-1}&=\quad\frac{48}{\de_2\de_3}\tau_3\tau_1(\tau_1-\de_1)\ , \non\\
&A_{0,1,-1}&=\quad-\frac{45\tau_3}{\de_2^2}\ , \non\\
&A_{0,2,-1}&=\quad\frac{45\tau_3}{2\de_2^2}\ , \non\\
&A_{0,-3,1}&=\quad\frac{256}{45\de_3^2}\tau_2(\tau_2-\de_2)(\tau_2-2\de_2)\ , \non\\
&A_{0,-3,2}&=\quad\frac{128}{45\de_3^2}\tau_2(\tau_2-\de_2)(\tau_2-2\de_2)\ . \non
\end{eqnarray}

\begin{figure}[!h]
\begin{center}
    \begin{picture}(300,275)
    \color{Gray}
    \put(0,29){\line(16,-1){250}}
    \put(0,89){\line(16,-1){250}}
    \put(0,149){\line(16,-1){250}}
    \put(0,209){\line(16,-1){250}}
    %
    \put(20,51){\line(16,-1){250}}
    \put(20,111){\line(16,-1){250}}
    \put(20,171){\line(16,-1){250}}
    \put(20,231){\line(16,-1){250}}
    %
    \put(40,73){\line(16,-1){250}}
    \put(40,133){\line(16,-1){250}}
    \put(40,193){\line(16,-1){250}}
    \put(40,253){\line(16,-1){250}}
    %
    \put(60,95){\line(16,-1){250}}
    \put(60,155){\line(16,-1){250}}
    \put(60,215){\line(16,-1){250}}
    \put(60,275){\line(16,-1){250}}
    %
    \put(250,13.2){\line(10,11){60}}
    \put(250,73.2){\line(10,11){60}}
    \put(250,133.2){\line(10,11){60}}
    \put(250,193.2){\line(10,11){60}}
    \put(200,16.3){\line(10,11){60}}
    \put(200,76.3){\line(10,11){60}}
    \put(200,136.3){\line(10,11){60}}
    \put(200,196.3){\line(10,11){60}}
    \put(150,19.5){\line(10,11){60}}
    \put(150,79.5){\line(10,11){60}}
    \put(150,139.5){\line(10,11){60}}
    \put(150,199.5){\line(10,11){60}}
    \put(100,22.7){\line(10,11){60}}
    \put(100,82.7){\line(10,11){60}}
    \put(100,142.7){\line(10,11){60}}
    \put(100,202.7){\line(10,11){60}}
    \put(50,25.8){\line(10,11){60}}
    \put(50,85.8){\line(10,11){60}}
    \put(50,145.8){\line(10,11){60}}
    \put(50,205.8){\line(10,11){60}}
    \put(0,29){\line(10,11){60}}
    \put(0,89){\line(10,11){60}}
    \put(0,149){\line(10,11){60}}
    \put(0,209){\line(10,11){60}}
    \color{red} \put(164,96){\Huge $\bullet$}
    \color{black}
    \put(140,187){\circle*{10}}
    \put(90,190){\circle*{10}}
    \put(260,143){\circle*{10}}
    \put(190,124){\circle*{10}}
    \put(210,146){\circle*{10}}
    \put(140,126){\circle*{10}}
    \put(270,35){\circle*{10}}
    \put(160,148){\circle*{10}}
    \put(90,129){\circle*{10}}
    \put(260,82){\circle*{10}}
    \put(140,66){\circle*{10}}
    \put(210,85){\circle*{10}}
    \put(170,42){\circle*{10}}
    \put(150,20){\circle*{10}}
    \put(20,111){\circle*{10}}
    \put(220,38){\circle*{10}}
    \put(100,82){\circle*{10}}
    \put(120,104){\circle*{10}}
    \put(150,80){\circle*{10}}
    \put(20,171){\circle*{10}}
    \put(20,231){\circle*{10}}
    \put(-15,86){$0$}
    \put(-19,29){$-\delta_3$}
    \put(-15,146){$\delta_3$}
    \put(-15,206){$2\delta_3$}
    \put(280,33){$0$}
    \put(290,48){$-\delta_1$}
    \put(260,13){$\delta_1$}
    \put(304,69){$-2\delta_1$}
    \put(145,7){$0$}
    \put(195,5){$\delta_2$}
    \put(237,2){$2\delta_2$}
    \put(85,11){$-\delta_2$}
    \put(38,13){$-2\delta_2$}
    \put(-9,15){$-3\delta_2$}
    \put(-20,116){\large $\boldsymbol{\tau_3}$}
    \put(120,0){\large $\boldsymbol{\tau_2}$}
    \put(295,25){\large $\boldsymbol{\tau_1}$}
    \end{picture}
\caption{\small Graphical representation of the 22-point discrete
operator (\ref{hdiscr}). The origin is shown in
red.}\label{Figure_disc}
\end{center}
\end{figure}
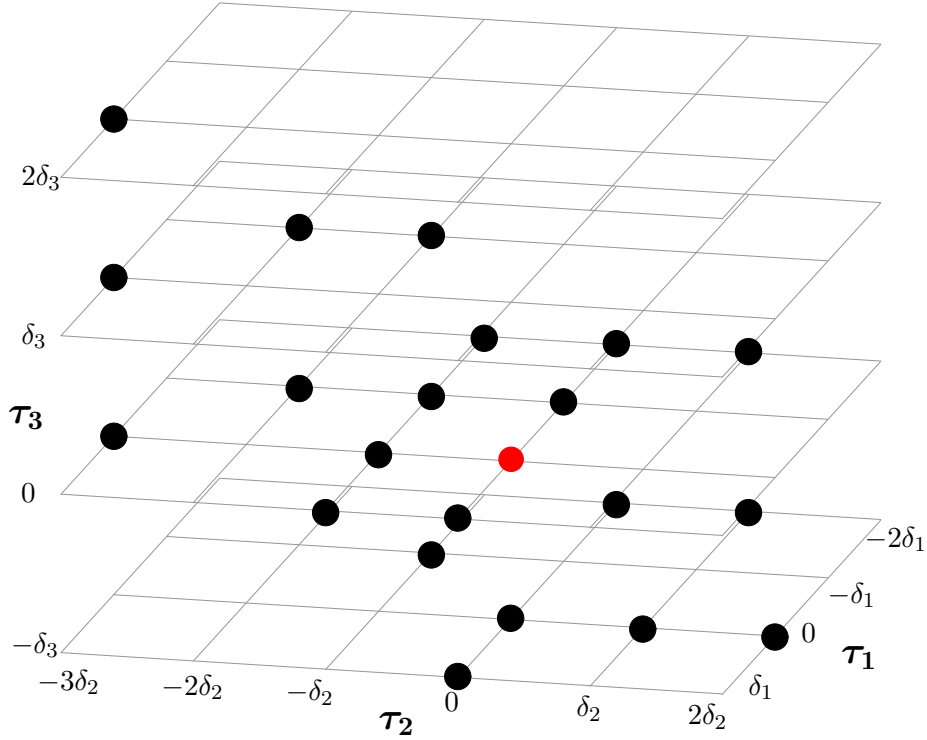

The corresponding eigenvalue problem is
\begin{equation}
 \sum_{k_1,k_2,k_3} A_{k_1 k_2 k_3}\
 \varphi(\tau_1+k_1\de_1,\tau_2+k_2\de_2,\tau_3+k_3\de_3)\ =\
 -2\epsilon\ \varphi(\tau_1,\tau_2,\tau_3)\ ,
\end{equation}
which is in the explicit form
\begin{equation}
\label{disc_eigen_ex}
\begin{aligned}
&\frac{2}{\de_1}\left[(2+\de_1\om)\left(\frac{\tau_1}{\de_1}+\frac{3\tau_2}
  {\de_2}+\frac{5\tau_3}{\de_3}\right)
  -3(1+10\nu)\right]\ \varphi(\tau_1,\tau_2,\tau_3)\\
&+\frac{2}{\de_1}\left[\frac{2\tau_1}{\de_1}+\frac{12\tau_2}{\de_2}
  +\frac{20\tau_3}{\de_3}+3(1+10\nu)\right]\varphi(\tau_1+\de_1,\tau_2,\tau_3)\\
&+\frac{4}{\de_1^2}(1+\de_1\om)\tau_1\
  \varphi(\tau_1-\de_1,\tau_2,\tau_3)+\frac{12}{\de_1\de_2}(2+\de_1\om)\tau_2\ \varphi(\tau_1,\tau_2-\de_2,\tau_3)\\
\end{aligned}
\end{equation}
\begin{equation*}
\begin{aligned}
&+\frac{48}{5\de_2}\tau_1(\tau_1-\de_1)\left[\frac{2\tau_2}{\de_2}
  +\frac{5\tau_3}{\de_3}+(1+5\nu)\right]\varphi(\tau_1-2\de_1,\tau_2,\tau_3)\\
&+\frac{5}{2}\left[\frac{8}{\de_1\de_3}(2+\de_1\om)+\frac{9}{\de_2^2}\right]\tau_3\
  \varphi(\tau_1,\tau_2,\tau_3-\de_3)
  -\frac{24\tau_2}{\de_1\de_2}\ \varphi(\tau_1+\de_1,\tau_2-\de_2,\tau_3)\\
&\frac{128}{45\de_3^2}\tau_2(\tau_2-\de_2)(\tau_2-2\de_2)\
   \varphi(\tau_1,\tau_2-3\de_2,\tau_3)
   -\frac{40\tau_3}{\de_1\de_3}\ \varphi(\tau_1+\de_1,\tau_2,\tau_3-\de_3)\\
&-\frac{32}{15\de_3}\tau_1\tau_2\left[\frac{\tau_2}{\de_2}
   -\frac{20\tau_3}{\de_3}-5(1+2\nu)\right]\varphi(\tau_1-\de_1,\tau_2-\de_2,\tau_3)\\
&+\frac{32}{15\de_2\de_3}\tau_1\tau_2(\tau_2-\de_2)\ \varphi(\tau_1-\de_1,\tau_2-2\de_2,\tau_3)\\
&+\frac{32}{15\de_3}\tau_1\tau_2\left[\frac{\tau_2}{\de_2}
    -\frac{10\tau_3}{\de_3}-5(1+2\nu)\right]\varphi(\tau_1-\de_1,\tau_2
    -\de_2,\tau_3+\de_3)\\
&-\frac{64}{3\de_3^2}\tau_1\tau_2\tau_3\
    \varphi(\tau_1-\de_1,\tau_2-\de_2,\tau_3-\de_3)\\
&-\frac{32}{15\de_2\de_3}\tau_1\tau_2(\tau_2-\de_2)\
    \varphi(\tau_1-\de_1,\tau_2-2\de_2,\tau_3+\de_3)\\
&-\frac{48}{5\de_2}\tau_1(\tau_1-\de_1)\left[\frac{\tau_2}{\de_2}
    +\frac{5\tau_3}{\de_3}+(1+5\nu)\right]\varphi(\tau_1-2\de_1,\tau_2+\de_2,\tau_3)\\
&-\frac{32}{15\de_2\de_3}\tau_1\tau_2(\tau_2-\de_2)\
    \varphi(\tau_1-\de_1,\tau_2-2\de_2,\tau_3+\de_3)\\
&-\frac{48}{5\de_2}\tau_1(\tau_1-\de_1)\left[\frac{\tau_2}{\de_2}
    +\frac{5\tau_3}{\de_3}+(1+5\nu)\right]\varphi(\tau_1-2\de_1,\tau_2+\de_2,\tau_3)\\
&-\frac{48}{5\de_2^2}\tau_2\tau_1(\tau_1-\de_1)\
    \varphi(\tau_1-2\de_1,\tau_2-\de_2,\tau_3)\\
&-\frac{48}{\de_2\de_3}\tau_3\tau_1(\tau_1-\de_1)\
    \varphi(\tau_1-2\de_1,\tau_2,\tau_3-\de_3)\\
&+\frac{48}{\de_2\de_3}\tau_3\tau_1(\tau_1-\de_1)\
    \varphi(\tau_1-2\de_1,\tau_2+\de_2,\tau_3-\de_3)\\
&-\frac{45\tau_3}{\de_2^2}\
    \varphi(\tau_1,\tau_2+\de_2,\tau_3-\de_3)+\frac{45\tau_3}{2\de_2^2}\ \varphi(\tau_1,\tau_2+2\de_2,\tau_3-\de_3)\\
&+\frac{256}{45\de_3^2}\tau_2(\tau_2-\de_2)(\tau_2-2\de_2)\
    \varphi(\tau_1,\tau_2-3\de_2,\tau_3+\de_3)\\
&+\frac{128}{45\de_3^2}\tau_2(\tau_2-\de_2)(\tau_2-2\de_2)\
    \varphi(\tau_1,\tau_2-3\de_2,\tau_3+2\de_3) = -2\epsilon\
    \varphi(\tau_1,\tau_2,\tau_3)\ .
\end{aligned}
\end{equation*}
It defines the discrete uniform $H_3$ system. It is worth noting
that, although we started from a second-order differential operator,
the 22-point finite-difference operator occurs -- it connects the
function in 22 different points in the lattice space: four points in
$\tau_1$-direction, six points in $\tau_2$-direction, four points in
$\tau_3$-direction. The structure of the operator is shown in
Fig.\ref{Figure_disc}.

The spectrum of the discrete operator $\tilde{h}_{H_3}$ (\ref{hdiscr})
for polynomial eigenfunctions coincides to the spectrum
of the continuous operator $h_{H_3}$ (\ref{h_H3_tau}) where all
eigenfunctions are polynomial ones:
\begin{equation}
\epsilon_{n_1,n_2,n_3}=2\om(n_1+3n_2+5n_3)\ ,
\end{equation}
where the $n$'s are non-negative integers. The eigenfunctions of
(\ref{hdiscr}) are related to the eigenfunctions of the continuous
operator $h_{H_3}$ by replacing each monomial with a quasi-monomial
in each variable. Such a phenomenon can be called {\it partial} or
{\it polynomial isospectrality}. We cannot exclude the existence of
other eigenstates of the discrete operator $\tilde{h}_{H_3}$ than
those given by polynomial eigenfunctions. These eigenstates can
correspond to non-normalizable eigenfunctions of $h_{H_3}$.

As a particular case let us consider the unit spacing
$\de_1=\de_2=\de_3=1$. This case corresponds to the discretization
in a cubic lattice the space of orbits (in $\tau$-space) with unit lattice vector.
The equation (\ref{disc_eigen_ex}) is reduced to the equation
\begin{equation*}
\begin{aligned}
&2\left[(2+\om)\left(\tau_1+3\tau_2+5\tau_3\right)-3(1+10\nu)\right]\ \varphi(\tau_1,\tau_2,\tau_3)\\
&+2\left[2\tau_1+12\tau_2+20\tau_3+3(1+10\nu)\right]\varphi(\tau_1+1,\tau_2,\tau_3)
\end{aligned}
\end{equation*}
\begin{equation}
\begin{aligned}
&+4(1+\om)\tau_1\ \varphi(\tau_1-1,\tau_2,\tau_3)+12(2+\om)\tau_2\ \varphi(\tau_1,\tau_2-1,\tau_3)\\
&+\frac{5}{2}(25+8\om)\tau_3\ \varphi(\tau_1,\tau_2,\tau_3-1)
-24\tau_2\ \varphi(\tau_1+1,\tau_2-1,\tau_3)\\
&+\frac{48}{5}\tau_1(\tau_1-1)\left[2\tau_2+5\tau_3+(1+5\nu)\right]\varphi(\tau_1-2,\tau_2,\tau_3)\\
&\frac{128}{45}\tau_2(\tau_2-1)(\tau_2-2)\
\varphi(\tau_1,\tau_2-3,\tau_3)-40\tau_3\ \varphi(\tau_1+1,\tau_2,\tau_3-1)\\
&-\frac{32}{15}\tau_1\tau_2\left[\tau_2-20\tau_3-5(1+2\nu)\right]\varphi(\tau_1-1,\tau_2-1,\tau_3)\\
&+\frac{32}{15}\tau_1\tau_2(\tau_2-1)\ \varphi(\tau_1-1,\tau_2-2,\tau_3)-\frac{48}{5}\tau_2\tau_1(\tau_1-1)\ \varphi(\tau_1-2,\tau_2-1,\tau_3)\\
&+\frac{32}{15}\tau_1\tau_2\left[\tau_2-10\tau_3-5(1+2\nu)\right]\varphi(\tau_1-1,\tau_2-1,\tau_3+1)\\
&-\frac{64}{3}\tau_1\tau_2\tau_3\ \varphi(\tau_1-1,\tau_2-1,\tau_3-1)-\frac{32}{15}\tau_1\tau_2(\tau_2-1)\ \varphi(\tau_1-1,\tau_2-2,\tau_3+1)\\
&-\frac{48}{5}\tau_1(\tau_1-1)\left[\tau_2+5\tau_3+(1+5\nu)\right]\varphi(\tau_1-2,\tau_2+1,\tau_3)\\
&-48\tau_3\tau_1(\tau_1-1)\ \varphi(\tau_1-2,\tau_2,\tau_3-1)+48\tau_3\tau_1(\tau_1-1)\ \varphi(\tau_1-2,\tau_2+1,\tau_3-1)\\
&-45\tau_3\ \varphi(\tau_1,\tau_2+1,\tau_3-1)+\frac{45\tau_3}{2}\ \varphi(\tau_1,\tau_2+2,\tau_3-1)\\
&+\frac{256}{45}\tau_2(\tau_2-1)(\tau_2-2)\ \varphi(\tau_1,\tau_2-3,\tau_3+1)+\frac{128}{45}\tau_2(\tau_2-1)(\tau_2-2)\
\varphi(\tau_1,\tau_2-3,\tau_3+2)\\
&+\frac{128}{45}\tau_2(\tau_2-1)(\tau_2-2)\
\varphi(\tau_1,\tau_2-3,\tau_3+2) = -2\epsilon\
\varphi(\tau_1,\tau_2,\tau_3)\ .
\end{aligned}
\end{equation}

A similar procedure of discretization can be applied to the integral
${\cal F}$. Instead of the continuous algebraic operator $f$ we get
its discrete counterpart
\begin{equation}
\label{fdiscr}
\tilde{f}\equiv
  f(\mathcal{D}_i^{(\de_i)},\mathcal{X}_i^{(\de_i)})\ =\
  \sum_{k_1,k_2,k_3} B_{k_1 k_2 k_3}\
  e^{k_1\de_1\pa_1+k_2\de_2\pa_2+k_3\de_3\pa_3}\ ,
\end{equation}
with the following coefficients
\begin{eqnarray}
&B_{0,-1,0}&=\ -\frac{3\tau_2}{\de_2}\left[\frac{10\tau_3}{\de_3}+\frac{12\tau_2}{\de_2}+5+30\nu\right]\ , \non\\
&B_{0,-2,0}&=\ \frac{18}{\de_2^2}\tau_2(\tau_2-\de_2)\ , \non\\
&B_{0,0,-1}&=\ -\frac{5\tau_3}{\de_3}\left[\frac{6\tau_2}{\de_2}
    +\frac{20\tau_3}{\de_3}-9+30\nu\right]\ , \non\\
&B_{0,0,-2}&=\ \frac{50}{\de_3^2}\tau_3(\tau_3-\de_3)\ , \non\\
&B_{-1,1,0}&=\ \frac{12\tau_3}{\de_2\de_3}\tau_1(\tau_1-\de_1)(\tau_1-2\de_1)\ , \non\\
\end{eqnarray}
\begin{eqnarray}
&B_{-1,1,-1}&=\ \frac{3}{2} \frac{\tau_1\tau_3}{\de_2}\left[\frac{8}{\de_3}(\tau_1-\de_1)(\tau_1-2\de_1)+\frac{15}{\de_2}\right]\ , \non\\
&B_{-1,2,-1}&=\ -\frac{45}{4\de_2^2}\tau_1\tau_3\ , \non\\
&B_{-1,0,-1}&=\ -\frac{3}{4\de_3}\tau_1\tau_3\left[ \frac{15}{\de_3}-\frac{16}{\de_2}(\tau_1-\de_1)(\tau_1-2\de_1)\right]\ , \non \\
&B_{-1,-3,0}&=\ -\frac{64}{45\de_3^2}\tau_1\tau_2(\tau_2-\de_2)(\tau_2-2\de_2)\ , \non\\
&B_{-1,-3,1}&=\ \frac{128}{45\de_3^2}\tau_1\tau_2(\tau_2-\de_2)(\tau_2-2\de_2)\ , \non\\
&B_{-1,-3,2}&=\ -\frac{64}{45\de_3^2}\tau_1\tau_2(\tau_2-\de_2)(\tau_2-2\de_2)\ , \non\\
&B_{-2,-1,0}&=\
\frac{8}{15\de_3}\tau_1(\tau_1-\de_1)\tau_2\left[\frac{\tau_2}{\de_2}
    -\frac{40\tau_3}{\de_3}-9-20\nu\right]\ , \non\\
&B_{-2,-2,0}&=\ -\frac{8}{15\de_2\de_3}\tau_1(\tau_1-\de_1)\tau_2(\tau_2-\de_2)\ , \non\\
&B_{-2,-1,1}&=\ -\frac{8}{15\de_3}\tau_1(\tau_1-\de_1)\tau_2
    \left[\frac{\tau_2}{\de_2}-\frac{20\tau_3}{\de_3}-9-20\nu\right]\ , \non\\
&B_{-2,-1,-1}&=\ \frac{32}{3\de_3^2}\tau_1(\tau_1-\de_1)\tau_2\tau_3\ , \non\\
&B_{-2,-2,1}&=\ \frac{8}{15\de_2\de_3}\tau_1(\tau_1-\de_1)\tau_2(\tau_2-\de_2)\ , \non\\
&B_{-3,1,0}&=\ \frac{24}{5\de_2}\tau_1(\tau_1-\de_1)(\tau_1-2\de_1)\left[\frac{\tau_2}{\de_2}+1+5\nu\right]\ , \non \\
&B_{-3,-1,0}&=\ \frac{24}{5\de_2^2}\tau_1(\tau_1-\de_1)(\tau_1-2\de_1)\tau_2\ , \non\\
&B_{0,-1,-1}&=\ \frac{30}{\de_2\de_3}\tau_2\tau_3\ . \non\\
\non
\end{eqnarray}

\begin{figure}[!h]
\begin{center}
    \begin{picture}(300,340)
    \color{Gray}
    \put(0,29){\line(16,-1){250}}
    \put(0,89){\line(16,-1){250}}
    \put(0,149){\line(16,-1){250}}
    \put(0,209){\line(16,-1){250}}
    \put(0,269){\line(16,-1){250}}
    \put(20,51){\line(16,-1){250}}
    \put(20,111){\line(16,-1){250}}
    \put(20,171){\line(16,-1){250}}
    \put(20,231){\line(16,-1){250}}
    \put(20,291){\line(16,-1){250}}
    \put(40,73){\line(16,-1){250}}
    \put(40,133){\line(16,-1){250}}
    \put(40,193){\line(16,-1){250}}
    \put(40,253){\line(16,-1){250}}
    \put(40,313){\line(16,-1){250}}
    \put(60,95){\line(16,-1){250}}
    \put(60,155){\line(16,-1){250}}
    \put(60,215){\line(16,-1){250}}
    \put(60,275){\line(16,-1){250}}
    \put(60,335){\line(16,-1){250}}
    \put(250,13.2){\line(10,11){60}}
    \put(250,73.2){\line(10,11){60}}
    \put(250,133.2){\line(10,11){60}}
    \put(250,193.2){\line(10,11){60}}
    \put(250,253.2){\line(10,11){60}}
    \put(200,16.3){\line(10,11){60}}
    \put(200,76.3){\line(10,11){60}}
    \put(200,136.3){\line(10,11){60}}
    \put(200,196.3){\line(10,11){60}}
    \put(200,256.3){\line(10,11){60}}
    \put(150,19.5){\line(10,11){60}}
    \put(150,79.5){\line(10,11){60}}
    \put(150,139.5){\line(10,11){60}}
    \put(150,199.5){\line(10,11){60}}
    \put(150,259.5){\line(10,11){60}}
    \put(100,22.7){\line(10,11){60}}
    \put(100,82.7){\line(10,11){60}}
    \put(100,142.7){\line(10,11){60}}
    \put(100,202.7){\line(10,11){60}}
    \put(100,262.7){\line(10,11){60}}
    \put(50,25.8){\line(10,11){60}}
    \put(50,85.8){\line(10,11){60}}
    \put(50,145.8){\line(10,11){60}}
    \put(50,205.8){\line(10,11){60}}
    \put(50,265.8){\line(10,11){60}}
    \put(0,29){\line(10,11){60}}
    \put(0,89){\line(10,11){60}}
    \put(0,149){\line(10,11){60}}
    \put(0,209){\line(10,11){60}}
    \put(0,269){\line(10,11){60}}
    \color{red} \put(144,134){\Huge $\bullet$}
    \color{black}
    \put(169,102){\circle*{10}}
    \put(140,187){\circle*{10}}
    \put(90,190){\circle*{10}}
    \put(140,126){\circle*{10}}
    \put(150,20){\circle*{10}}
    \put(100,82){\circle*{10}}
    \put(150,80){\circle*{10}}
    \put(20,171){\circle*{10}}
    \put(20,231){\circle*{10}}
    \put(170,162){\circle*{10}}
    \put(210,206){\circle*{10}}
    \put(101,143){\circle*{10}}
    \put(51,146){\circle*{10}}
    \put(219,159){\circle*{10}}
    \put(270,96){\circle*{10}}
    \put(220,99){\circle*{10}}
    \put(20,291){\circle*{10}}
    \put(140,247){\circle*{10}}
    \put(90,250){\circle*{10}}
    \put(260,203){\circle*{10}}
    \put(160,209){\circle*{10}}
    \put(-15,86){$-\delta_3$}
    \put(-19,29){$-2\delta_3$}
    \put(-15,146){$0$}
    \put(-15,206){$\delta_3$}
    \put(-15,266){$2\delta_3$}
    \put(280,33){$-\delta_1$}
    \put(290,48){$-2\delta_1$}
    \put(260,13){$0$}
    \put(304,69){$-3\delta_1$}
    \put(145,7){$0$}
    \put(195,5){$\delta_2$}
    \put(237,2){$2\delta_2$}
    \put(85,11){$-\delta_2$}
    \put(38,13){$-2\delta_2$}
    \put(-9,15){$-3\delta_2$}
    \put(-25,116){\large $\boldsymbol{\tau_3}$}
    \put(120,0){\large $\boldsymbol{\tau_2}$}
    \put(295,20){\large $\boldsymbol{\tau_1}$}
    \end{picture}
\caption{\small Graphical representation of the 22-point discrete
operator (\ref{fdiscr}). The origin is shown in
red.}
\label{Figure_disc_int}
\end{center}
\end{figure}
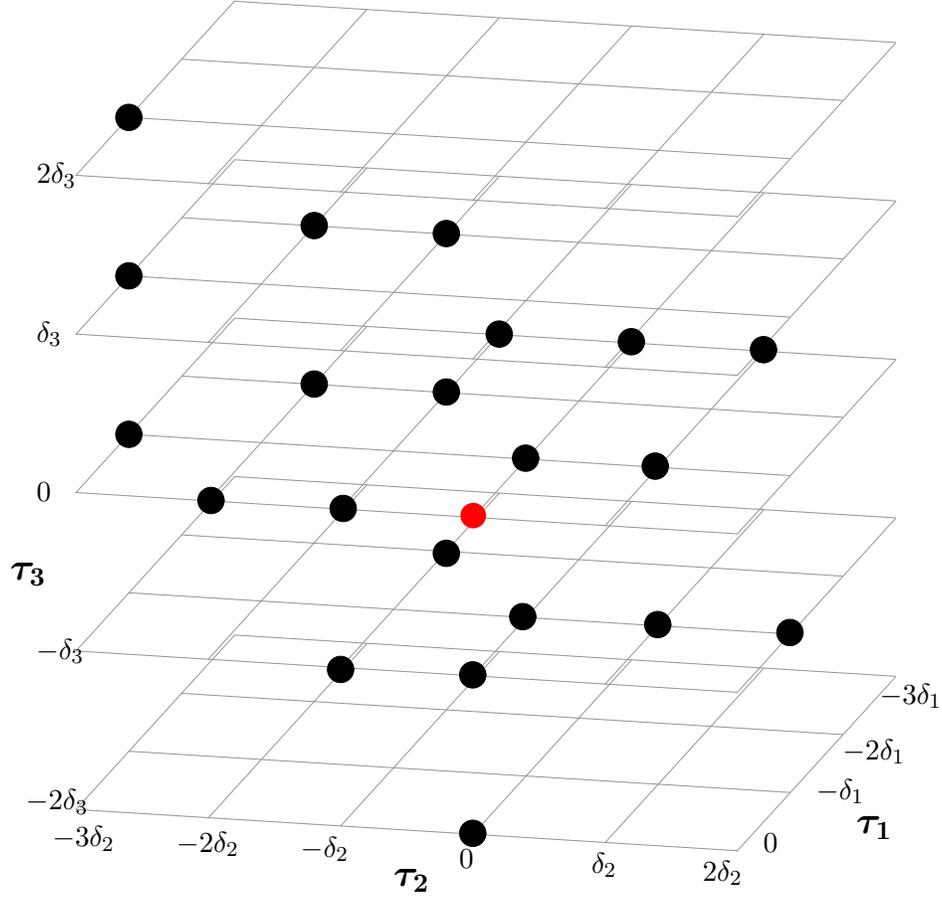
We end up with the 22-point finite-difference operator occurs -- it
connects the function in 22 different points in the lattice space:
four points in $\tau_1$-direction, six points in $\tau_2$-direction,
five points in $\tau_3$-direction. The structure of the operator is
shown in Fig.\ref{Figure_disc_int}. It is surprising that both
discrete operators $\tilde h_{H_3}$ and $\tilde f$ have the same
structure connecting 22 points.


It is evident that the discrete operators $\tilde{h}_{H_3}$ and $\tilde f$ continue to commute. Such a procedure of discretization preserves the property of integrability.

\medskip

\section{Quasi-Exactly-Solvable generalization}

Among the eigenfunctions of the Hamiltonian $h_{\rm H_3}$ (\ref{h_H3_tau}) there is
an infinite family of eigenfunctions depending on the single variable $\tau_1$.
These eigenstates are solutions of the eigenvalue problem
\begin{equation}
\label{h_H3_tau1}
-h_1\ \vphi\equiv -4\tau_1\frac{\pa^2\vphi}{\pa\tau_1^2}
+(4\om\tau_1-6(1+10\nu))\frac{\pa\vphi}{\pa\tau_1}\ =\ \ep\vphi\ .
\end{equation}
Corresponding eigenfunctions are given by the Laguerre polynomials and the eigenvalues are linear in quantum number
\begin{equation}
\label{laguerre}
\vphi_{n_1}(\tau_1)\ =\ L_{n_1}^{(1+30\nu)/2}(\om\tau_1)\ ,\quad
\epsilon_{n_1}=4\om n_1\ ,\quad n_1=0,1,2,\ldots
\end{equation}
The operator in the l.h.s. of (\ref{h_H3_tau1}) can be rewritten in
terms of the generators $J_k^0,J^-$ of the Cartan subalgebra of the
algebra $sl(2)$ of the first order differential operators:
\begin{equation}
\label{sl2}
J_k^+=\tau_1^2\frac{\pa}{\pa\tau_1}-k\tau_1\ , \quad
J_k^0=\tau_1\frac{\pa}{\pa\tau_1}-\frac{k}{2}\ , \quad
J^-=\frac{\partial}{\pa\tau_1}\ ,
\end{equation}
(see \cite{Turbiner:2005_2}). For integer $k$ the generators
(\ref{sl2}) have a common invariant subspace in polynomials
of degree not higher than $k$,
\begin{equation}
\label{pol_space}
\mathcal{P}_k=\langle \tau_1^p\ | \ 0\leq p\leq k\rangle\ ,
\end{equation}
where $\dim \mathcal{P}_k=(k+1)$. The operator (\ref{h_H3_tau1})
takes the $sl(2)$-Lie-algebraic form
\begin{equation}
\label{h_sl2}
h_1=4J_0^0 J^- - 4\om J_0^0+6(1+10\nu)J^-\ .
\end{equation}
It is easy to check that the operator $h_1$ preserves an infinite
flag of spaces of polynomials (\ref{pol_space}),
\begin{equation}
\label{flag}
\mathcal{P}_0\subset\mathcal{P}_1\subset\mathcal{P}_2\subset\cdots
\subset\mathcal{P}_k\subset\cdots\ ,
\end{equation}
and, in particular, any eigenfunction is an element of the flag.

Let us proceed to construct a QES generalization of (\ref{H_H3}). We
look for the QES Hamiltonian in a certain form
\begin{equation}
\label{H_qes}
H^{(qes)}_{H_3}=H_{H_3}+V^{(qes)}(\tau_1)\ ,
\end{equation}
where $V^{(qes)}$ is a potential. Let us make a gauge rotation of
(\ref{H_qes}) of the form (\ref{h_H3}). We impose the requirement
that the resulting operator possesses a $\tau_1$-depending family of
eigenfunctions. We obtain the following equation:
\begin{equation}
\label{h_H3_tau1_qes}
-h_1^{(qes)}\vphi\ \equiv \
 -4\tau_1\frac{\partial^2\vphi}{\partial\tau_1^2}
 +(4\om\tau_1-6(1+10\nu))\frac{\partial\vphi}{\partial\tau_1}
+2V^{(qes)}(\tau_1)\vphi\ =\ \epsilon\vphi\ .
\end{equation}

Our aim is to find $V^{(qes)}$ for which the operator $h_1^{(qes)}$
is $sl(2)$-Lie-algebraic -- can be rewritten in terms of the
generators (\ref{sl2}). Following \cite{Turbiner:2005_2}, let us
gauge rotate the operator (\ref{h_H3_tau1_qes}),
\begin{equation}
\begin{aligned}
 h_1^{(sl(2)-qes)}=&\ \tau_1^{-\gamma}\exp\left(\frac{a}{4}\tau_1^2\right)h_1^{(qes)}
 \tau_1^{\gamma}\exp\left(-\frac{a}{4}\tau_1^2\right)\\
 =&\ 4\tau_1\frac{\pa^2}{\pa\tau_1^2}
 -2(2a\tau_1^2+2\om\tau_1-3-4\gamma-30\nu)\frac{\pa}{\pa\tau_1}
 +a^2\tau_1^3+2a\om\tau_1^2\\
 &-2a\left(2\gamma+15\nu+\frac{5}{2}\right)\tau_1+
 \frac{2\gamma(2\gamma+30\nu+1)}{\tau_1}-4\om\gamma-2V^{(qes)}(\tau_1)\ .
 \end{aligned}
\end{equation}
If the corresponding potential $V^{(qes)}$ is chosen of the form
\begin{equation}
 V^{(qes)}=\frac{1}{2}\ a^2\tau_1^3+a\om\tau_1^2-a\left(2k+2\gamma + 15\nu+
 \frac{5}{2}\right)\tau_1 +
 \frac{\gamma(2\gamma+30\nu+1)}{\tau_1}\ ,
\end{equation}
the operator $h_1^{(sl(2)-qes)}$ has the Lie-algebraic form
\begin{equation}
\label{h_sl2_qes}
 h_1^{(sl(2)-qes)}\ =\ 4J_k^0 J^--4aJ_k^+ -4\om
 J_k^0+2(k+4\gamma+3(1+10\nu))J^- \ ,
\end{equation}
where $a \geq 0$ and $\gamma$ are parameters, here the constant terms
are dropped off.

It can be seen that the operator $h_1^{(sl(2)-qes)}$ (see (\ref{h_sl2_qes}))
has the space $\mathcal{P}_k$ as the invariant subspace, but it does not preserve the flag of spaces (\ref{flag}). Hence, (\ref{h_sl2}) has (k+1) polynomial eigenfunctions of the form of polynomials of the degree $k$,
\[
 P_j^{(k)}(\tau_1)=\sum_{i=0}^k \gamma_i^{(j)}\tau_1^i\ ,\quad
 j=0,1,2,\ldots\ ,
\]
while other eigenfunctions are not polynomials. Now we can give the final expression of the $sl(2)$-quasi-exactly-solvable Hamiltonian associated with the root
space $H_3$:
\begin{equation}
\label{H_H3_qes}
\begin{aligned}
    H_{H_3}^{(qes)}\ = \ &\frac{1}{2}\sum_{k=1}^{3}\left[-\frac{\pa^{2}}{\pa
    x_{k}^{2}}+\om^{2}x_{k}^{2}+\frac{g}{x_{k}^{2}}\right]\\
    &+\sum_{\{i,j,k\}}\,\sum_{\mu_{1,2}=0,1}\frac{2g}{[x_{i}+(-1)^{\mu_1}
    \varphi_{+}x_{j}+(-1)^{\mu_2}\varphi_{-}x_{k}]^{2}}\\
    &+\frac{1}{2}\ a^2(\mathbf{x}^2)^3+a\om(\mathbf{x}^2)^2-a\left(2k+2\gamma
    +15\nu+\frac{5}{2}\right)\mathbf{x}^2\\
    &+\frac{\gamma(2\gamma+30\nu+1)}{\mathbf{x}^2}\ ,
\end{aligned}
\end{equation}
where $\{i,j,k\}=\{1,2,3\}$ and its even permutations, and
$\mathbf{x}^2=\sum_{i=1}^3x_i^2$. For this Hamiltonian we know
$(k+1)$ eigenstates explicitly. Their eigenfunctions are of the form
\begin{equation}
\Psi_k(x)\ =\ \Delta_{1}^{\nu}\Delta_{2}^{\nu}\
(\mathbf{x}^2)^{\gamma}\cdot P_k(\mathbf{x}^2)\
\mathrm{e}^{-\frac{\om}{2}\mathbf{x}^2-\frac{a}{4}(\mathbf{x}^2)^2}
\end{equation}
where $P_k$ is a polynomial of degree $k$, the coupling constant
$g=\nu(\nu-1)>-\frac{1}{4}$ and $\De_{1,2}$ are given by
(\ref{D12}). It is worth presenting several $P_k$ explicitly,
\begin{equation}
\begin{aligned}
P_0(\mathbf{x}^2)&=1\ ,\quad
E_0=\frac{3}{2}\om\left(1+10\nu+\frac{4}{3}\gamma\right)\ ,\\[15pt]
P_{1,\pm}(\mathbf{x}^2)&\ =\ \mathbf{x}^2+\frac{1}{2a}\left[\om\pm\sqrt{\om^2+2a(4\gamma+3(1+10\nu))}\right]\ ,\\
E_{1,\pm}&\ =\ E_0+\om\mp\sqrt{\om^2+2a(4\gamma+3(1+10\nu))}\ .
\end{aligned}
\end{equation}
The solutions for $k=1$ are related through analytic continuation in
one of the parameters $\om,a,\gamma,\nu$ keeping other parameters
fixed. They form two-sheeted Riemann surface.

The QES Hamiltonian (\ref{H_H3_qes}) is integrable -- the integral
${\cal F}$ (\ref{opf}) remains to commute with the Hamiltonian.

\medskip

\section{Hidden algebra}

We have shown that the Hamiltonian in the algebraic form
(\ref{h_H3_tau}) acts on the finite-dimensional spaces of
multivariate polynomials $\mathcal{P}_n^{(1,2,3)},\ n=0,1,2,\ldots$
(see (\ref{minflag})). A goal of this Section is to show that
each one of these subspaces is a representation space of an
infinite-dimensional algebra of differential operators which
we call $h^{(3)}$ and to study this algebra.

The algebra $h^{(3)}$ is infinite-dimensional but
finitely-generated. Their generating elements can be split into two
classes. The first class of generators (lowering and Cartan
operators) act in $\mathcal{P}^{(1,2,3)}_n$ for any $n \in N$ and
therefore they preserve the flag $\mathcal{P}^{(1,2,3)}$. The second
class operators (raising operators) act on the space
$\mathcal{P}^{(1,2,3)}_n$ for a certain value of $n$ only; they do
not act on a space at other $n$'s.

Let us introduce the following notation for the derivatives:
\[
\pa_i\equiv\frac{\pa}{\pa\tau_i}\ ,\quad
\pa_{ij}\equiv\frac{\pa^2}{\pa\tau_{i}\pa\tau_{j}}\
,\quad\pa_{ijk}\equiv\frac{\pa^3}{\pa\tau_{i}\pa\tau_{j}\pa\tau_{k}}\
.
\]
The first class of generating elements consist of the 22 generators where 13 of them are the first order operators
\begin{equation}
\begin{aligned}
\label{ops_1}
& T_0^{(1)}=\pa_1\,, && T_0^{(2)}=\pa_2\,, && T_0^{(3)}=\pa_3\,,\\
& T_1^{(1)}=\tau_1\pa_1\,, && T_2^{(2)}=\tau_2\pa_2\,, && T_3^{(3)}=\tau_3\pa_3\,,\\
& T_1^{(3)}=\tau_1\pa_3\,, && T_{11}^{(3)}=\tau_1^2\pa_3\,, && T_{111}^{(3)}=\tau_1^3\pa_3\,,\\
& T_1^{(2)}=\tau_1\pa_2\,, && T_{11}^{(2)}=\tau_1^2\pa_2\,, && T_2^{(3)}=\tau_2\pa_3\,,\\
& &&T_{12}^{(3)}=\tau_1\tau_2\pa_3\ ,&&
\end{aligned}
\end{equation}
the 6 are of the second order
\begin{equation}
\begin{aligned}
\label{ops_2}
& T_2^{(11)}=\tau_2\pa_{11}\,, && T_{22}^{(13)}=\tau_2^2\pa_{13}\,, && T_{222}^{(33)}=\tau_2^3\pa_{33}\,,\\
& T_3^{(12)}=\tau_3\pa_{12}\,, && T_3^{(22)}=\tau_3\pa_{22}\,, &&
T_{13}^{(22)}=\tau_1\tau_3\pa_{22}\ ,
\end{aligned}
\end{equation}
and 2 are of the third order
\begin{equation}
\begin{aligned}
\label{ops_3}
& T_3^{(111)}=\tau_3\pa_{111}\,, &&
T_{33}^{(222)}=\tau_3^2\pa_{222}\ .
\end{aligned}
\end{equation}

The generators of the second class consist of 8 operators where 1 of them is of the first order
\begin{equation}
\label{R1}
J_1^+ = \tau_1J_0\ ,
\end{equation}
4 are of the second order
\begin{equation}
\begin{aligned}
\label{R2}
& J_{2,-1}^+=\tau_2\pa_1J_0\,, &&
J_{3,-2}^+=\tau_3\pa_2J_0\,,  && J_{22,-3}^+ = \tau_2^2\pa_3J_0\,,
&& J_2^+ = \tau_2J_0(J_0+1)\ ,
\end{aligned}
\end{equation}
and 3 are of the third order
\begin{equation}
\begin{aligned}
\label{R3}
& J_{3,-11}^{+}=\tau_3\pa_{11}J_0\ , &&
J_{3,-1}^+=\tau_3\pa_1J_0(J_0+1)\ , &&
J_3^+=\tau_3J_0(J_0+1)(J_0+2)\ ,
\end{aligned}
\end{equation}
where we have introduced the diagonal operator
\begin{equation}
\label{jo}
J_0=\tau_1\pa_1+2\tau_2\pa_2+3\tau_3\pa_3-n\ .
\end{equation}
for a convenience. In fact, this operator is identity operator,
it is of the zeroth order and, hence, it belongs to the first class.

Before to proceed to study the commutation relations between generators
we introduce a notion of conjugation. Let $T_1$ and $T_2$ be operators
acting on a monomial. We say that $T_2$ is a conjugate to $T_1$ if the
operator $T_2 T_1$ leaves the monomial unchanged. There is a certain
ambiguity related with the central operator $J_0$ (\ref{jo}). Formally,
it seems self-conjugated. From another side, the operator $J_0$ is,
in fact, the unit operator. Thus, one can define a conjugate to $J_0$
to be equal to 1 and visa versa: a conjugate to 1 is equal $J_0$.
>From this point of view any operator is defined up to a multiplicative
factor of $J_0$. We resolve this ambiguity by defining the generators (\ref{ops_1})-(\ref{jo}) in a way for supporting below-presented commutation
relations and eventually a structure of algebra. In particular, $T_{0}^{(1)}$ is conjugated to $J_{1}^{+}$,
\[
    \pa_1 \longleftrightarrow \tau_1 J_0 \ ,
\]
and $T_{0}^{(3)}$ is conjugated to $J_{3}^{+}$,
\[
    \pa_3 \longleftrightarrow \tau_3 J_0 (J_0+1) (J_0+2)\ .
\]
Except for this ambiguity, the conjugation coincides with the Fourier transform.

A certain number of generating operators (\ref{ops_1})-(\ref{R3})
span ten Abelian subalgebras:
\begin{eqnarray}
\label{subals}
L=\{T_{0}^{(3)},T_{1}^{(3)},T_{11}^{(3)},T_{111}^{(3)}\}&
\longleftrightarrow
 &\mathfrak{L}=\{T_{3}^{(111)},J_{3,-11}^{+},J_{3,-1}^{+},J_{3}^{+}\} \non\\
R=\{T_{0}^{(2)},T_{1}^{(2)},T_{11}^{(2)}\}&
\longleftrightarrow
 &\mathfrak{R}=\{T_{2}^{(11)},J_{2,-1}^{+},J_{2}^{+}\} \non\\
F=\{T_{2}^{(3)},T_{12}^{(3)}\}&
\longleftrightarrow
 &\mathfrak{F}=\{T_{3}^{(12)},J_{3,-2}^{+}\}\\
E=\{T_{3}^{(22)},T_{13}^{(22)}\}&
\longleftrightarrow
 &\mathfrak{E}=\{T_{22}^{(13)},J_{22,-3}^{+}\} \non\\
G=\{T_{222}^{(33)}\}&\longleftrightarrow&\mathfrak{G}=\{T_{33}^{(222)}\}
  \non
\end{eqnarray}
The remaining generators span a non-commutative algebra isomorphic
to $gl(2)\oplus\mathcal{R}^{(2)}$, where $\mathcal{R}^{(2)}$ is the
Abelian algebra of dimension 2,
\begin{equation}
\label{subalb}
B=\{T_{0}^{(1)},T_{1}^{(1)},T_{2}^{(2)},T_{3}^{(3)},J_{0},J_{1}^{+}\}
\end{equation}
The arrow in (\ref{subals}) means that the corresponding operators are related by conjugation (for example, $T_0^{(3)}$ and $J_3^+$). The subalgebra $B$ is
the unique algebra those operators are self-conjugated.

Decomposition (\ref{subals}),(\ref{subalb}) allows us to give a
compact representation of the commutation relations relating the
generating elements. As first we can show that the commutators
between the Abelian algebras from the l.h.s. (r.h.s.) of
(\ref{subals}) are closed -- they are either zero or written in
terms of generators of one of the algebras from l.h.s. (r.h.s.) plus
from the algebra $B$
\begin{align*}
&[L,R]=0, && [\mathfrak{L},\mathfrak{R}]=0,\\
&[L,F]=0, && [\mathfrak{L},\mathfrak{F}]=0,\\
&[L,E]=P_{2}(R), && [\mathfrak{L},\mathfrak{E}]=P_{2}(\mathfrak{R}),\\
&[L,G]=0, && [\mathfrak{L},\mathfrak{G}]=0,\\
&[R,F]=L, && [\mathfrak{R},\mathfrak{F}]=\mathfrak{L},\\
&[R,E]=0, && [\mathfrak{R},\mathfrak{E}]=0,\\
&[R,G]=P_{2}(F), && [\mathfrak{R},\mathfrak{G}]=P_{2}(\mathfrak{F}),\\
&[F,E]=P_{2}(R\oplus B), && [\mathfrak{F},\mathfrak{E}]=P_{2}(\mathfrak{R}\oplus B),\\
&[F,G]=0, && [\mathfrak{F},\mathfrak{G}]=0,\\
&[E,G]=P_{3}(F\oplus B), &&
[\mathfrak{E},\mathfrak{G}]=P_{3}(\mathfrak{F}\oplus B),
\end{align*}
Here $P_k(Q)$ means that the commutator is given by a polynomial of the
$k$-th degree in the generators of $Q$. It turns out all commutators
are symmetric under conjugation. This property also holds for cross commutators
of the algebras from the l.h.s. and the r.h.s. in (\ref{subals}),
\begin{align*}
&[L,\mathfrak{R}]=P_{2}(F\oplus B), && [\mathfrak{L},R]=P_{2}(\mathfrak{F}\oplus B),\\
&[L,\mathfrak{F}]=P_{2}(R\oplus B), && [\mathfrak{L},F]=P_{2}(\mathfrak{R}\oplus B),\\
&[L,\mathfrak{E}]=P_{2}(F), && [\mathfrak{L},E]=P_{2}(\mathfrak{F}),\\
&[L,\mathfrak{G}]=P_{2}(R\oplus E), && [\mathfrak{L},G]=P_{2}(\mathfrak{R}\oplus\mathfrak{E}),\\
&[R,\mathfrak{F}]=E, && [\mathfrak{R},F]=\mathfrak{E},\\
&[R,\mathfrak{E}]=P_{2}(F\oplus B), && [\mathfrak{R},E]=P_{2}(\mathfrak{F}\oplus B),\\
&[R,\mathfrak{G}]=0, && [\mathfrak{R},G]=0,\\
&[F,\mathfrak{E}]=G, && [\mathfrak{F},E]=\mathfrak{G},\\
&[F,\mathfrak{G}]=P_{2}(E\oplus B), && [\mathfrak{F},G]=P_{2}(\mathfrak{E}\oplus B),\\
&[E,\mathfrak{G}]=0, && [\mathfrak{E},G]=0,
\end{align*}
The commutator between any Abelian algebra from (\ref{subals}) and the algebra $B$ is of the type of a semidirect product:
\begin{align*}
[L,B]=L\,, && [R,B]=R\,, && [F,B]=F\,, && [E,B]=E\,, && [G,B]=G\,,\\
[\mathfrak{L},B]=\mathfrak{L}\,, && [\mathfrak{R},B]=\mathfrak{R}\,,
&& [\mathfrak{F},B]=\mathfrak{F}\,, &&
[\mathfrak{E},B]=\mathfrak{E}\,, && [\mathfrak{G},B]=\mathfrak{G}\,,
\end{align*}
At last, let us indicate commutators between conjugated algebras in (\ref{subals}) - all of them are polynomials in generators of $B$:
\begin{align*}
&[L,\mathfrak{L}]=P_{3}(B), && [R,\mathfrak{R}]=P_{2}(B), && [F,\mathfrak{F}]=P_{2}(B),\\
&[E,\mathfrak{E}]=P_{3}(B), && [G,\mathfrak{G}]=P_{4}(B).\\
\end{align*}
Latter two types of relations are represented by triangular
diagrams, see for example, Fig.3.
\begin{figure}[!h]
\begin{center}
\begin{picture}(110,110)
\put(55,50){\vector(-1,0){45}} \put(55,50){\vector(1,0){45}}
\put(55,50){\vector(0,1){45}} \put(105,45){$\mathfrak{L}$}
\put(0,45){L} \put(43,35){$P_{3}(B)$} \put(50,100){$B$}
\put(75,80){\scalebox{1.7}{\rotatebox{315}{$\ltimes$}}}
\put(15,87){\scalebox{1.7}{\rotatebox{225}{$\ltimes$}}}
\end{picture}
\caption{\small Triangular diagram relating the subalgebras $L$,
$\mathfrak{L}$ and $B$. It is a generalization of Gauss
decomposition for semi-simple algebras.}
\label{Figure3}
\end{center}
\end{figure}
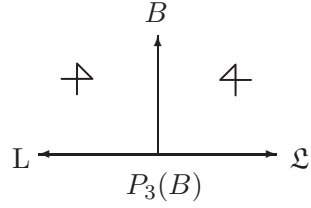
In general, the commutation relations between two operators are
characterized by a non-linear combination of the generators in
r.h.s.. Calculating double, triple, etc. commutators one can see
that the commutation relations can not be closed at any order and
the order of the r.h.s. is increasing. Hence, the $h^{(3)}$ algebra
is not a polynomial algebra. It is the infinite dimensional algebra
of ordered monomials in the 30 generating operators
(\ref{ops_1})-(\ref{R3}) shown above.

Since $h^{(3)}$ is the algebra of differential operators acting on
$\mathcal{P}^{(1,2,3)}_n$ it should be possible to write the
$h_{H3}$ Hamiltonian (\ref{h_H3_tau}) as a combination of the
(flag-preserving) generating elements (\ref{ops_1})-(\ref{ops_3}) of
$h^{(3)}$. The $h^{(3)}$-algebraic form of the $H_3$ model
(\ref{H_H3}) is the following:
\begin{equation}
\begin{aligned}
h_{H_3}=&\ 4T_1^{(1)}T_0^{(1)} + 24T_2^{(2)}T_0^{(1)} +
40T_3^{(3)}T_0^{(1)} - \frac{48}{5}T_2^{(2)}T_{11}^{(2)}
+\frac{45}{2}T_3^{(22)} + \frac{32}{15}T_{12}^{(3)}T_2^{(2)} \\
& - 48T_3^{(3)}T_{11}^{(2)} - \frac{64}{3}T_3^{(3)}T_{12}^{(3)} +
\frac{128}{45}T_{222}^{(33)} + (6+60\nu)T_0^{(1)} - 4\om
T_{1}^{(1)} \\
& - \frac{48}{5}(1+5\nu)T_{11}^{(2)} - 12\om T_{2}^{(2)} -
\frac{64}{15}(2+5\nu)T_{12}^{(3)} - 20\om T_{3}^{(3)}\ .
\end{aligned}
\end{equation}

\section{Conclusions}

We have shown that the $H_3$ rational system related to the
non-crystallographic root system $H_3$ is exactly solvable with the
characteristic vector $(1,2,3)$ \footnote{It is worth noting that in
the past there existed a single attempt to study the $H_3$ rational
system \cite{Ruehl:1998}}. This work complements the previous
studies of the rational (and trigonometric) models, related with
crystallographic root systems (e.g. \cite{Turbiner:1998_1} -
\cite{Turbiner:2009}). A certain significance of exploration of the
$H_3$ rational system is due to a fact that this model is defined in
three-dimensional Euclidian (physical) space. There are very few
known exactly-solvable systems in this space -- the Coulomb problem,
four-body Calogero-Sutherland $(A_3)$ and $BC_3$
rational-trigonometric models among them. Surprisingly, all of them
are integrable while the Coulomb, $A_3$ and $BC_3$ rational
problems are superintegrable. The same is correct for all known
two-dimensional exactly-solvable problems in the Euclidean space:
all of them are integrable.

Taking Coxeter invariants of $H_3$ as coordinates provided us a way
to reduce the rational $H_3$ Hamiltonian to algebraic form. It gave
us a chance to find the eigenfunctions of the rational $H_3$
Hamiltonian which are proportional to polynomials in these invariant
coordinates. It seems correct that these eigenfunctions exhaust all
eigenfunctions in the Hilbert space. It is worth noting that the
matrix $A_{ij}(\tau)$ which appears in front of the second
derivatives after changing variables in Laplacian from Cartesian to
the $H_3$ Coxeter invariant coordinates (see Eqs. (\ref{A-B})) has
polynomial entries corresponding to flat space metric, hence the
Riemann tensor vanishes. We call metric the {\it Arnold} metric
\footnote{Many years ago V.I.~Arnold \cite{Arnold:1976} pointed out
that the contravariant flat metric on the space of orbits of any
Coxeter group written in terms of the polynomial invariants has
polynomial matrix elements. We have to add that for such a metric
coefficient functions in front of the first derivative terms in the
Laplace-Beltrami operator are also polynomials
\cite{Turbiner:2005_1}}.

It should be stressed that it was stated that the Hamiltonian of the
$H_3$ rational system (\ref{H}) is completely integrable
\cite{Sasaki:2000}. This implies the existence of two
mutually-commuted operators (the `higher Hamiltonians') which
commute with the Hamiltonian forming a commutative algebra. It is
known (see \cite{Olshanetsky:1983}) for the crystallographic systems
that these higher Hamiltonians are the differential operators of the
degrees which coincide to the minimal degrees of the root space (the
Lie algebra) or their squares for the $A_N$ case. It may suggest that
for the $H_3$ rational system the commuting integrals might be
differential operators of the orders six and ten. Their explicit forms
are not known so far. It is evident that these commuting operators
should take on an algebraic form after a gauge rotation (with the
ground state function as a gauge factor), and a change of variables
from Cartesian coordinates to the Coxeter invariant variables $\tau$'s.
Following the experience with different integrable systems, it seems the
integral(s) related with separation of variables do not enter to the
commutative algebra. Therefore, the integral ${\cal F}$ is out of
the commutative algebra of integrals. It might serve as an
indication to a superintegrability of the $H_3$ rational system.

An analysis similar to the analysis of this paper has not yet been
presented for the case of the rational systems related to the
non-crystallographic root spaces $H_4$. A study in progress indicates
that the characteristic vector for the quantum $H_4$ integrable system
is $(1,5,8,12)$. In the case of dihedral group $I_2(k)$ the rational model has
the characteristic vector $(1,k)$ \cite{TTW:2009}.
It should be pointed out that unlike the rational models
it is not possible to construct integrable (and exactly-solvable)
trigonometric systems related to the non-crystallographic root spaces
as a natural generalization of the Hamiltonian Reduction Method \cite{Olshanetsky:1983}.

The existence of algebraic form of the $H_3$ rational
Olshanetsky-Perelomov Hamiltonian makes possible the study of their
polynomial perturbations which are invariant wrt the $H_3$ Coxeter
group by purely algebraic means: one can develop a perturbation
theory in which all corrections are found by linear algebra methods
\cite{Tur-pert}. In particular, it gives a chance to calculate the $H_3$
Coxeter-invariant, polynomial correlation functions by algebraic means.

Another important property of the existence of algebraic form of the
$H_3$ rational Hamiltonian is a chance to perform a canonical,
Lie-algebraic discretization to uniform (see Ch.IV) and exponential
\cite{Chriss:2001} lattices. In the case of both lattices such a
discretization preserves a property of integrability, polynomiality
of the eigenfunctions remains and it is isospectral. Although it does not
give a hint how to introduce a scalar product for a discrete model.
Making the weighted projective transformation (\ref{wpt}) of the
$H_3$ algebraic form (\ref{h_H3_tau}) we arrive at different
algebraic form of the $H_3$ Hamiltonian. Making then the
Lie-algebraic discretization we arrive at a discrete model related
to an original discrete model via change of variables. It can be
considered as a definition of a polynomial change of variables for
discrete operators.

We found the $sl(2)$-quasi-exactly-solvable generalization of the
$H_3$ model which remains integrable. This is the first example of
quasi-exact-solvability related to non-crystallographic root
systems. It complements the results obtained previously for all
rational models related to crystallographic systems (see
\cite{Turbiner:2005_2}) and for the $I_2(k)$ rational model
\cite{TTW:2009} -- each of these models admit a certain $sl(2)$-QES
generalization.

Thank to the explicit knowledge of the ground state function (\ref{Psi_H3})
supersymmetric $H_3$ model can be constructed following a procedure
realized in \cite{Freedman:1990} for $A_N$ rational model, in
\cite{Brink:1998} for the $BC_N$ rational model and in \cite{Quesne:2010}
for the $I_2(k)$ rational model. It can be done elsewhere.

\bigskip

\textit{\small Acknowledgements}. The computations in this paper
were performed on MAPLE 8 and MAPLE 11 with the packages COXETER and
WEYL created by J.~Stembridge.

\end{document}